\documentclass[12pt]{article}
\usepackage[centertags]{amsmath}
\usepackage{amsfonts}
\usepackage{epsfig}
\usepackage{amssymb}
\usepackage{amsthm}
\usepackage{newlfont}
\usepackage{amstext}

\theoremstyle{plain}
\newtheorem{Th}{Theorem}[section]
\newtheorem{Cor}[Th]{Corollary}
\newtheorem{Lem}[Th]{Lemma}
\newtheorem{Prop}[Th]{Proposition}

\theoremstyle{definition}
\newtheorem{Def}{Definition}[section]
\newtheorem{Ex}{Example}[section]
\theoremstyle{remark}
\newtheorem*{Rem}{Remark}
\numberwithin{equation}{section}
\newcommand{\AAf}{{\mathbb A}}
\newcommand{\CC}{{\mathbb C}}
\newcommand{\EE}{{\mathbb E}}

\newcommand{\II}{{\mathbb I}}

\newcommand{\PP}{{\mathbb P}}
\newcommand{\RR}{{\mathbb R}}

\newcommand{\ZZ}{{\mathbb Z}}

\newcommand{\cP}{{\mathcal P}}

\newcommand{\calC}{{\mathcal C}}

\newcommand{\ba}{{\boldsymbol a}}
\newcommand{\bC}{{\boldsymbol C}}
\newcommand{\bob}{{\boldsymbol b}}
\newcommand{\boe}{{\boldsymbol e}}
\newcommand{\vbe}{\vec{\boe}}
\newcommand{\bx}{{\boldsymbol x}}

\newcommand{\vbx}{\vec{\bx}}

\newcommand{\bu}{{\boldsymbol u}}

\newcommand{\by}{{\boldsymbol y}}
\newcommand{\vby}{\vec{\by}}

\newcommand{\bv}{{\boldsymbol v}}

\newcommand{\vbv}{\vec{\bv}}

\newcommand{\bX}{{\boldsymbol X}}
\newcommand{\bP}{{\boldsymbol P}}

\newcommand{\bY}{{\boldsymbol Y}}
\newcommand{\bOm}{{\boldsymbol \Omega}}
\newcommand{\bzero}{{\boldsymbol 0}}

\newcommand{\tv}{\tilde{v}}
\newcommand{\tbX}{\tilde{\bX}}
\newcommand{\tbY}{\tilde{\bY}}

\newcommand{\tA}{\tilde{A}}
\newcommand{\tH}{\tilde{H}}

\newcommand{\tQ}{\tilde{Q}}

\newcommand{\tF}{\tilde{F}}
\newcommand{\tP}{\tilde{P}}

\newcommand{\D}{{\Delta}}
\newcommand{\tD}{\tilde{\Delta}}
\newcommand{\pp}{{\partial}}
\newcommand{\ra}{\rightarrow}

\begin{document}

\begin{center}
{\Large \sc The symmetric, $D$-invariant and Egorov reductions of the
quadrilateral lattice}
\bigskip

{\large Adam Doliwa$^{1,\dagger}$ and Paolo Maria Santini$^{2,3,\S}$}

\bigskip

{\it $^1$Instytut Fizyki Teoretycznej, Uniwersytet Warszawski \\
ul. Ho\.{z}a 69, 00-681 Warszawa, Poland

\smallskip

$^2$Istituto Nazionale di Fisica Nucleare, Sezione di Roma\\
P.le Aldo Moro 2, I--00185 Roma, Italy

\smallskip

$^3$Dipartimento di Fisica, Universit\`a di Roma "La Sapienza"\\
P.le Aldo Moro 2, I--00185 Roma, Italy
}

\bigskip

$^\dagger$e-mail: {\tt Adam.Doliwa@fuw.edu.pl}

$^{\S}$e-mail:  {\tt Paolo.Santini@roma1.infn.it}

\bigskip

{}

\end{center}


\begin{abstract}

\noindent We present a detailed study of the geometric and algebraic
properties of the
multidimensional quadrilateral lattice (a lattice whose elementary
quadrilaterals are planar; the discrete analogue of a conjugate
net) and of its basic reductions. To make this study, 
we introduce the
notions of forward and backward data, which allow us to give a
geometric meaning to the $\tau$--function of the lattice, defined
as the potential connecting these data. 
Together with the known circular lattice (a lattice whose
elementary quadrilaterals can be inscribed in circles; the
discrete analogue of an orthogonal conjugate net) 
we introduce and study two other basic and independent reductions
of the quadrilateral lattice: the symmetric lattice, for which the forward
and backward data coincide, and the $d$-invariant lattice, characterized
by the invariance of a certain natural frame along the main diagonal.
We finally discuss the Egorov lattice, which is, at the same time,
symmetric, circular and $d$-invariant. The integrability
properties of all these lattices are established using geometric,
algebraic and analytic means;  in particular we present a
$\bar\partial$ formalism to construct large classes of such
lattices. We also discuss quadrilateral hyperplane lattices and
the interplay between quadrilateral point and hyperplane lattices
in all the above reductions. 

\medskip

{\it 1998 PACS:} 04.60.Nc, 02.40.Hw

\end{abstract}

\newpage
\section{Introduction}

In a recent paper~\cite{MQL} we have introduced the notion of
"Multidimensional Quadrilateral Lattice" (MQL), i.e., a lattice
$\bx : \ZZ^N \rightarrow \PP^M$, $N\leq M$, with all its
elementary quadrilaterals planar, which is the discrete analogue
of a multidimensional conjugate net~\cite{DarbouxOS}. Furthermore,
we showed that the planarity constraint (which is a linear
constraint) provides a way to construct the lattice uniquely, once
a suitable set of initial data is given. 

In this paper we  present a detailed study of three
basic and independent integrable reductions of the
quadrilateral lattice: the symmetric lattice, the circular lattice and the
$d$-invariant lattice; we also study the
Egorov lattice which is, at the same time, symmetric, circular and
$d$-invariant.
All these reductions satisfy additional geometric properties which
are compatible with the planarity constraint of the MQL.

The {\it symmetric} lattice follows from the observation that one can
associate, with a given quadrilateral lattice, forward and backward data
connected through a potential coinciding with the $\tau$--function of the
lattice, and it corresponds to the particular situation
in which the backward and forward rotation coefficients 
coincide. The {\it circular} lattice,
discrete analogue of an orthogonal net, is instead characterized
by the fact that all its elementary quadrilaterals are inscribed
in circles. The {\it $d$-invariant} lattice is a MQL characterized by the
invariance of a certain natural frame along the main diagonal.
The {\it Egorov} lattice, discrete analogue of a Egorov
net~\cite{DarbouxOS,Bianchi}, is simultaneously symmetric, 
circular and $d$-invariant (for $N=M$), and can be equivalently 
characterized by the 
fact that a
pair of opposite angles of the elementary quadrilateral consists
of right angles. 

The geometric properties characterizing the above reductions 
make use of the connections between point lattices and hyperplane
lattices (lattices in the dual space $(\PP^M)^*$). In some cases the 
connection comes from additional structure in
the ambient space $\PP^M$; in some other cases, it is a consequence of 
the inner
symmetry of the lattice. The precise connections between point and
hyperplane lattices corresponding to all the above reductions
are also presented in this paper.

Our presentation reflects the effort of constructing a general theory of 
the
MQL and of its reductions and therefore the results will not appear in a
chronological order of derivation but rather in a logical order.

Although the research field of integrable discrete geometry is relatively
new, the amount of associated results is already very large and it
is often difficult to go through the corresponding
literature, also because many of these results
are not even published, having being presented only during conferences or
seminars, or private conversations. A brief but hopefuly correct account of
the literature closed to the subject considered in this paper is the 
following.

The proper discrete
analogue of a conjugate net on a surface was first proposed by
Sauer~\cite{Sauer}. The MQL equations were first derived by Bogdanov and
Konopelchenko~\cite{BoKo} as integrable discrete analogues of the Darboux
equations for conjugate nets, but without any geometric
characterization.
The notion of circular lattice was first proposed by Martin, de Pont and
Scharrock~\cite{2dcl1} and
Nutborne~\cite{2dcl2}
for $N=2$, $M=3$, as a discrete analogue of surfaces parametrized
by curvature lines (see also~\cite{BP2}); later by Bobenko for
$N=M=3$ \cite{Bobenko} and, finally, for arbitrary $N\leq M$ by
Cie\'{s}li\'{n}ski, Doliwa and Santini \cite{CDS}; subsequently
Konopelchenko and Schief have shown~\cite{KoSchief2} that circular 
lattices in $\EE^3$ can be
conveniently characterized by solutions of the $2+1$ dimensional discrete
sine-Gordon equation~\cite{NiSchief}.
A geometric proof of the integrability of the circular
lattice
was first given in~\cite{CDS} while the analytic proof of its
integrability was given in~\cite{DMS} through the $\bar\partial$
method.
The notion of Egorov lattice with its right angles characterization was
found by Schief~\cite{Schief-priv1}. In the derivation of the Egorov lattice,
he apparently used the algebraic formulation of the symmetric constraint; 
this formulation was restricted 
to the subclass of circular lattices and its geometric
meaning was not given~\cite{Schief-priv2}. He also found the $d$-invariance
of the Egorov lattice (the Killing vector property)~\cite{Schief-priv3}.
The finite-gap formulations of the circular and Egorov lattices have
also recently appeared in the literature~\cite{Krich-priv}.

The new results written down in this paper, although already presented in 
several occasions~\cite{conferences}, are the following:\\
1) The geometric meaning of the $\tau$-function of the MQL;\\
2) The theory of integrable hyperplane lattices, and its central role in the
reduction theory of MQL;\\
3) The algebraic and geometric notions of symmetric and $d$-invariant
lattices, as basic and independent reductions of the MQL;\\
4) The successful application of the $\bar\pp$ reduction method, already
used in the case of circular lattices~\cite{DMS}, to all the other
reductions. 

After this work was completed and presented we were told that, according
to Schief~\cite{Schief-priv4}, the algebraic definition of a symmetric 
lattice was
presented in~\cite{Schief-priv1}.

In the rest of this Introduction we summarize the basic
results on quadrilateral lattices and the known facts on
hyperplanes in projective spaces which will be used in the paper.

In Section~\ref{sec:backward} we introduce the "backward"
representation of the quadrilateral lattice and we show that
the compatibility between the backward construction and the
standard forward construction leads to the existence of a potential
which can be identified with the $\tau$-function
of the lattice. 

In Section~\ref{sec:h-lat} we first introduce the notion of quadrilateral
hyperplane lattice; then we introduce and study the notions of 
dual, adjoint, conjugate and complementary systems of point and hyperplane
lattices. 

In Section~\ref{sec:symm-latt} we study the first integrable
reduction, the symmetric lattice together with its integrability properties.

In Section~\ref{sec:circ-latt}
we discuss, in the same spirit, the second basic reduction, the circular
lattice. 

In Section~\ref{sec:d-latt} we define the third basic reduction, the
$d$-invariant lattice and study its properties. 

Section~\ref{sec:Eg-latt} is devoted to the study of the Egorov
lattice which is, at the same time, symmetric, circular and $d$-invariant. 

In Section~\ref{sec:D-bar} we finally study the integrability
properties of all the above lattices from the point of view of their
solvability, making use of a $\bar\partial$--reduction method recently
introduced in~\cite{ZakMa2} in the continuous case and generalized
in~\cite{DMS} to a discrete context.

We finally remark that the equations characterizing the above
lattices are potentially relevant also in physics, being integrable
discretizations of equations arising
in hydrodynamics~\cite{Dubr,GrundlandZelazny,Fer,Tsarev} and in quantum
field theory~\cite{Dubr-ass,DMMMS,Kyoto}.

\subsection{Quadrilateral point lattices}

\label{sec:for-ql}

\begin{figure}
\begin{center}
\epsffile{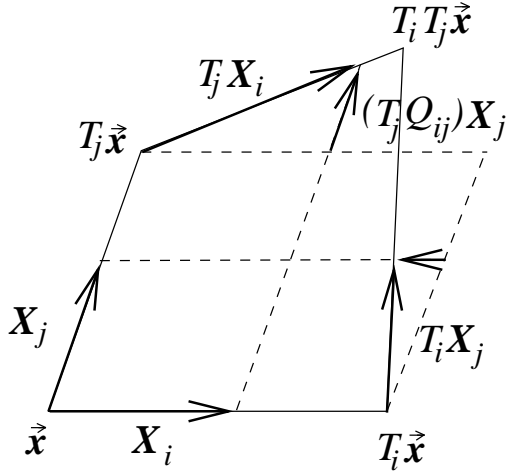}
\end{center}
\caption{Definition of the forward data}
\label{fig:forward}
\end{figure}

Consider a multidimensional quadrilateral lattice; i. e., a mapping
$x :\ZZ^N \ra \PP^M$, $N\leq M$,
with all the elementary quadrilaterals planar \cite{MQL}. In the affine
representation (in which the lattice is a mapping
$\vbx: \ZZ^N \rightarrow \RR^M$) the planarity
condition can be formulated in terms of the Laplace equations
\begin{equation}  \label{eq:Laplace}
\D_i\D_j\vbx=(T_{i} A_{ij})\D_i\vbx+
(T_j A_{ji})\D_j\vbx,\;\; i\not= j, \; \; \;  i,j=1 ,\dots, N,
\end{equation}
where $T_i$ is the translation operator in the $i$ direction, $\D_i = T_i -
1$ and the coefficients $A_{ij}$ satisfy the MQL equation
\begin{equation} \label{eq:MQL-A}
\D_k A_{ij} =
 (T_jA_{jk})A_{ij} +(T_k A_{kj})A_{ik} - (T_kA_{ij})A_{ik},
\;\; i\neq j\neq k\neq i.
\end{equation}
It is often convenient to reformulate  equations (\ref{eq:Laplace}) as a
first order system~\cite{MQL}. We introduce the suitably scaled tangent 
vectors $\bX_i$, $i=1,...,N$,
\begin{equation}  \label{def:HX}
\D_i\vbx = (T_iH_i) \bX_i,
\end{equation}
in such a way that the $j$-th variation of $\bX_i$
is proportional to $\bX_j$ only (see Figure~\ref{fig:forward})
\begin{equation} \label{eq:lin-X}
\D_j\bX_i = (T_j Q_{ij})\bX_j,    \; \; \; i\ne j \; .
\end{equation}
The compatibility condition for the system~(\ref{eq:lin-X})
gives the following new form of the MQL equations
\begin{equation} \label{eq:MQL-Q}
\D_kQ_{ij} = (T_kQ_{ik})Q_{kj}, \;\;\; i\neq j\neq k\neq i.
\end{equation}
The scaling factors $H_i$,
called the Lam\'e coefficients, solve the linear equations
\begin{equation} \label{eq:lin-H}
\D_iH_j = (T_iH_i) Q_{ij}, \; \; \; i\ne j \; ,
\end{equation}
 whose compatibility gives equations
(\ref{eq:MQL-Q}) again; moreover
\begin{equation}   \label{def:A-H}
A_{ij}= \frac{\D_j H_i}{H_i} \; , \; \; i\ne j \; .
\end{equation}
In~\cite{MQL} it was proven that, given $N(N-1)/2$ initial
quadrilateral surfaces, the quadrilateral lattice $\bx$
follows uniquely from the planarity constraint. To construct the initial surfaces,
one gives $N$ arbitrary intersecting initial curves $\vbx^{(0)}_i$,
$i=1,\dots , N$; the initial $(ij)$-surface  is then built uniquely
assigning the initial data $A_{ij}^{(0)}$, $i\neq j$, as functions of
$n_i$, $n_j$, via equation (\ref{eq:Laplace}). 
Equivalently, together with the $N$ intersecting initial curves,
we can give the initial data $\{ H_i^{(0)}, Q_{ij}^{(0)} \}$, meaning
that we give the coefficients $H_i^{(0)}$ (or, equivalently, the tangent vectors
$\bX_i^{(0)}$) on the $i$-th initial curve and then the data  $Q_{ij}^{(0)}$,
$i\ne j$, as functions of $n_i$, $n_j$. Therefore the solution of the MQL
equations depends on $N(N-1)$ arbitrary functions of two variables.

\begin{Rem}
To make the construction of the lattice possible, in our
considerations we assume we deal with {\it generic} lattices,
i.e., that the point $x$ and its nearest neighbours $T_1 x$,\dots,
$T_N x$ are in general position; in consequence, the subspace
$\langle x, T_1 x$,\dots, $T_N x \rangle $ is a linear subspace of
$\PP^M$ of maximal possible dimension $N$.
\end{Rem}

In this paper we study some distinguished reductions of the MQL
which posses additional geometric properties that, once imposed
on the initial surfaces "propagate" everywhere through the
construction of the lattice. Since the quadrilateral lattice
is integrable, these reductions will inherit its integrability
properties.

In the continuous limit: 
\begin{align} \label{eq:limit-x}
\D_i\vbx & \sim \varepsilon \frac{\partial}{\partial u_i } =
 \varepsilon \partial_i \; , \qquad 0 < \varepsilon < \! < 1 \; , \\
Q_{ij} & \sim \varepsilon\beta_{ij} \; , \label{eq:limit-Q}
\end{align}
the MQL reduces to an $N$ dimensional conjugate net in $\RR^M$, 
characterized by
the Darboux equations~\cite{DarbouxOS}
\begin{equation} \label{eq:Darboux}
\partial_k \beta_{ij} = \beta_{ik}\beta_{kj}, \quad i\ne j \ne k \ne i.
\end{equation}

\subsection{Hyperplanes}
\label{sec:hyp}
In Section~\ref{sec:h-lat} we introduce and study the properties of lattices
in the dual space, i.e., of hyperplane lattices.
These considerations will turn out to be relevant in the reduction theory
of the quadrilateral lattices, when the introduction of additional 
geometric structure will
allow to establish a direct connection between point lattices and 
hyperplane lattices.

To make the paper self contained, in the rest of this introduction
we summarize some basic known facts on the algebraic representation of 
the
projective space and of its dual.

Points of $\PP^M$ are directions (one dimensional linear subspaces)
of $\RR^{M+1}$ and they can be represented (up to
multiplication by a non-zero scalar factor) by non-zero vectors
of $\RR^{M+1}$. In a fixed basis $\boe_0, \boe_1,\dots ,\boe_M$ of 
$\RR^{M+1}$,
the coordinates
$\bu = (u^0,u^1 ,\dots u^M)^T$ of such
a vector are called the homogeneous coordinates of the
corresponding point $u=[\bu]$ of the projective space.

The hyperplanes of $\PP^M$ are $M$ dimensional linear subspaces
of $\RR^{M+1}$ and they can be represented (up to
multiplication by a non-zero scalar factor) by non-zero co-vectors
of $(\RR^{M+1})^*\equiv \RR^{M+1}$. The coordinates
$\ba^* = (a^*_0,a^*_1 ,\dots a^*_M)$ of such a
co-vector are called the homogeneous coordinates of the
corresponding hyperplane $a^*=[\ba^*]$, and the condition that the point
with homogeneous coordinates
$\bu = (u^0,u^1, \dots , u^M )^T $ belongs to
the hyperplane represented by $\ba^*=(a_0^*, a_1^* , \dots , a_M^*)$
is given by the linear homogeneous equation
\begin{equation} \label{eq:lin-hyp}
\langle \ba^* | \bu \rangle = a_0^* u^0 + a_1^* u^1 + \dots + a_M^* u^M = 0 \; .
\end{equation}

\begin{Rem}[Duality principle] Notice that equation~(\ref{eq:lin-hyp}) is
``symmetric" in the sense that the expression ``the point $u$
belongs to the hyperplane $a^*$" can be changed into ``the
hyperplane $a^*$ contains the point $u$". Geometrically, all
hyperplanes (points of $(\PP^M)^*$) passing through a fixed point
of $\PP^M$ form a hyperplane in $(\PP^M)^*$, which is represented
by this point; therefore $((\PP^M)^*)^* = \PP^M$.
\end{Rem}

By fixing a hyperplane $\PP^{M-1}_\infty$ in $\PP^M$, called then the
hyperplane at
infinity, we can represent the remaining (affine) part $\AAf^M =
\PP^M \setminus
\PP^{M-1}_\infty$ of the projective space by points $\vec{\bv}\in\RR^M$;
if the hyperplane at infinity
is characterized by $u^0=0$, then the points of the affine space can be
normalized
to $(1, u^1, \dots , u^M )^T$, and $\vec{\bu}=
( u^1, \dots , u^M )^T$.

Hyperplanes in
$\AAf^M$ can be represented (again, up to a non-zero factor) by
non-homogeneous linear
equations as follows
\begin{equation}
a_0^*  + a_1^* x^1 + \dots + a_M^* x^M = 0 \; .
\end{equation}
The representation can be made unique, by
affinization of $(\PP^M)^*$, i.e., by removing from $(\PP^M)^*$
hyperplanes passing through a fixed point of $\PP^M$. For our purposes we
assume
that this point belongs to $\AAf^M$, and we identify it with the 
origin of $\RR^M$.
Then the equation of any hyperplane which does not pass through the origin,
i.e.,
$a_0^*\ne 0$, can be normalized to have $a_0^* = - 1$. Such a 
hyperplane can
be represented by the co-vector $\vec{\ba^*}\in(\RR^M)^*$ and
consists of points $\vec{\bx}$ satisfying the equation
\begin{equation} \label{eq:h-af}
\langle \vec{\ba^*} | \vec{\bx} \rangle  = a_1^* x^1 + 
\dots + a_M^* x^M  = 1
\; .
\end{equation}

If $\vec{\bv}$ is a point of the hyperplane represented by
$\vec{\ba}^*$, then the parallel (in the standard sense)
hyperplane passing through $t\vec{\bv}$ is
represented
by $t^{-1}\vec{\ba}^*$; equivalently, the equation
of such a hyperplane can be written as
$\langle \vec{\ba}^* | \vbx \rangle = t$.
Taking the limit $t\to \infty$ we infer that the hyperplane at infinity
$\PP^{M-1}_\infty$
is represented by the zero co-vector $\vec{\bzero^*}$.
On the other hand, all the hyperplanes passing through 
$\vec\bzero \in\RR^M$
are
represented by "infinite" co-vectors; equivalently, the equation
of the hyperplane passing through $\vec\bzero$ and
parallel to that represented by $\vec{\ba^*}$
can be written as $\langle \vec{\ba^*} | \vec{\bx} \rangle = 0$.

Given two hyperplanes $a^*$ and $b^*$ represented by the co-vectors 
$\vec{\ba^*}$
and $\vec{\bob^*}$, the equation of the unique hyperplane passing through 
the
origin and containing their intersection $a^* \cap b^*$ is
\begin{equation}
\langle \vec{\ba^*} - \vec{\bob^*} | \vec{\bx} \rangle = 0 \; .
\end{equation}
\begin{Def} \label{def:co-parallelity}
Two subspaces of co-dimension $2$ are called "co-parallel" if there exists 
a
hyperplane passing through $\vec{\bzero}$ and containing them.
\end{Def}
\begin{Rem}
The above notion is dual to the parallelism of two lines in the affine space.
\end{Rem}
\begin{Cor}
Two co-dimension $2$ subspaces obtained by intersection of two pairs
of hyperplanes $a^*_i\cap b^*_i$, $i=1,2$, are co-parallel if the
corresponding co-vectors $\vec{\ba^*}_i - \vec{\bob^*}_i$, $i=1,2$ are
proportional.
\end{Cor}

A {\it correlation} is a projective mapping between a projective
space and its dual
\begin{equation*}
\calC: \PP^M \to (\PP^M)^* \; .
\end{equation*}
In the homogeneous description, such a mapping is given by a linear mapping
(given uniquely up to a non-zero scalar factor) between the vector
space $\RR^{M+1}$ and its dual; if $a^*=[\ba^*]$, $v=[\bv]$, and 
$a^* = \calC (v)$,
then the correlation $\calC$ is represented by a matrix $\bC$ such that
$\ba^* = (\bC\bv)^T$.

Any correlation $\calC$ defines its adjoint correlation
\begin{equation*}
\calC^*:((\PP^M)^*)^*=\PP^M \to (\PP^M)^*
\end{equation*}
being represented by the matrix $\bC^T$ transposed of $\bC$.
An important class of correlations is provided by {\it involutory
correlations}, that is correlations identical to their adjoints. 
Matrices of such
correlations must satisfy the condition the $\bC^T = \pm \bC$.

When the matrix of the correlation is symmetric, then the correlation 
is called
{\it polarity}; we denote it by $\cP$.
The image $\cP(v)$ of a point $v=[\bv]\in\PP^M$
is called the polar hyperplane of $v$; it
consists of points $x=[\bx]$ satisfying
equation $\langle \bP \bv | \bx \rangle = 0$ .

Any polarization $\cP$ defines the corresponding
quadric hypersurface $Q_\cP$, which consists of points
belonging to their polar hyperplanes: $x\in \cP(x)$;
in the homogeneous description, the quadric is given by
equation $\langle \bP \bx | \bx \rangle = 0 $.

\begin{Ex}
Consider the polarization whose quadric is the standard
sphere of radius $1$ centered at the origin:
\begin{equation*}
Q_\cP = S^{M-1}=\{ \vbx \in \EE^M | \vbx \cdot \vbx = 1 \} \; .
\end{equation*}
Then the polar hyperplane of a point $\vbv$ is the hyperplane
orthogonal to $\vbv$ and passing through the point $\vbv/(\vbv \cdot \vbv)$
(see Figure~\ref{fig:ex-pol}).
The polar of the origin is the hyperplane at infinity, therefore
this polarization is an affine mapping, i.e., it maps
parallel lines into co-parallel subspaces (of co-dimension two).
\begin{figure}
\begin{center}
\epsffile{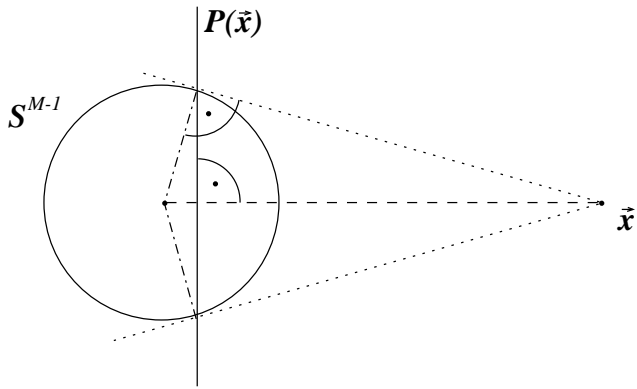}
\end{center}
\caption{Polarity with respect to a sphere}
\label{fig:ex-pol}
\end{figure}
\end{Ex}

\section{The backward representation of the quadrilateral lattice}

\label{sec:backward}

In this Section we define the backward data $\tbX_i$, $\tH_i$, $\tQ_{ij}$
of the quadrilateral lattice. It turns out that the relation between the
standard forward data $\bX_i$, $H_i$, $Q_{ij}$ and the backward data
is given in terms of the $\tau$--function, which is one of central
objects of the soliton theory.

\begin{figure}
\begin{center}
\epsffile{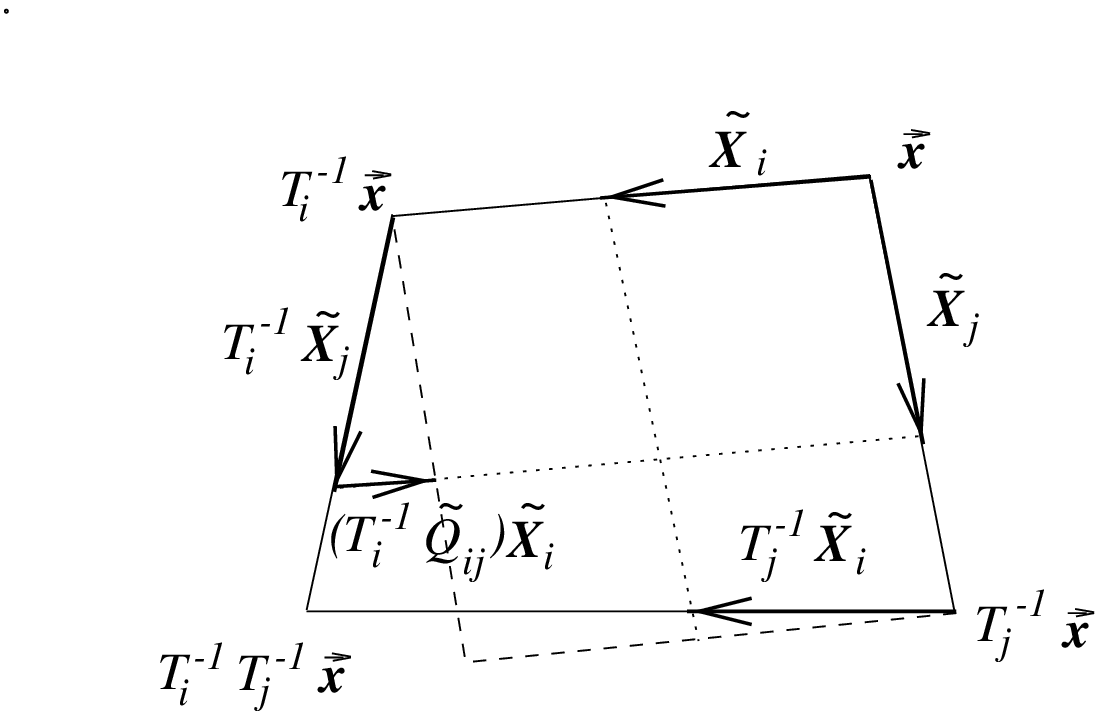}
\end{center}
\caption{Definition of the backward data}
\label{fig:back}
\end{figure}

The backward tangent vectors $\tbX_i$ and the backward Lam\'e coefficients
$\tH_i$, $i=1,\dots,N$ are defined with the help of the backward
difference operator $\tD_i := 1- T_i^{-1}$:
\begin{equation} \label{eq:b-H-X}
\tD_i\vbx = (T_i^{-1}\tH_i ) \tbX_i \; , \qquad \text{or}
\quad \D_i\vbx = \tH_i (T_i\tbX_i ) \, ;
\end{equation}
the backward Lam\'e coefficients
are again chosen in such a way (see Figure~\ref{fig:back}) that the
$\tD_i$ variation of $\tbX_j$ is proportional to $\tbX_i$ only. We
define the backward rotation coefficients $\tQ_{ij}$ as the
corresponding proportionality factors
\begin{equation} \label{eq:lin-bX}
\tD_i\tbX_j = (T_i^{-1} \tQ_{ij})\tbX_i \; , \qquad \text{or}
\quad \D_i\tbX_j =  (T_i\tbX_i)\tQ_{ij},    \quad i\ne j \; .
\end{equation}

Comparing equations (\ref{eq:lin-H}) and (\ref{eq:lin-bX})
we see immediately that the new functions $\tQ_{ij}$ satisfy the MQL 
equations~(\ref{eq:MQL-Q}) as well. Moreover the new scaling factors 
$\tH_i$ satisfy the following system of linear equations
\begin{equation} \label{eq:lin-bH}
\D_j\tH_i =  (T_j\tQ_{ij})\tH_{j},   \; \; \; i\ne j \; ,
\end{equation}
whose compatibility condition gives again the MQL equations 
(\ref{eq:MQL-Q}).

An easy consequence of equations \eqref{eq:b-H-X}, \eqref{eq:lin-bX},
\eqref{eq:lin-bH} is the following, obvious from a
geometric point of view, observation
\begin{Prop} \label{prop:b-f}
The vector function $\vbx: \ZZ^N \rightarrow \RR^M$ representing a
quadrilateral lattice satisfies the backward Laplace equation
\begin{equation} \label{eq:Laplace-b}
\tD_i\tD_j\vbx = (T_i^{-1}\tA_{ij})\tD_i\vbx +
(T_j^{-1}\tA_{ji})\tD_j\vbx
\; , \quad i\ne j \; ,
\end{equation}
where, in the notation of this Section,
\begin{equation}
\tA_{ij} = \frac{\tD_j \tH_i}{\tH_i}\; .
\end{equation}
\end{Prop}

The forward and backward rotation coefficients
$Q_{ij}$ and $\tQ_{ij}$ describe the same lattice $\vbx$
from different points of view, therefore one can expect their 
interrelation.
Indeed, defining the functions $\rho_i:\ZZ^N\to \RR$ as the 
proportionality factors between
$\bX_i$ and $T_i\tbX_i$ (both vectors are proportional to $\D_i\vbx$):
\begin{equation} \label{eq:def-rho}
\bX_i = - \rho_i ( T_i\tbX_i) \; , \qquad T_iH_i = - 
\frac{1}{\rho_i}\tH_i \; , \quad
i=1,\dots ,N \; ,
\end{equation}
we have the following

\begin{Prop} \label{prop:bf-data}
The forward and backward data of the lattice $\vbx$ are related 
through the following formulas
\begin{equation} \label{eq:Q-Qt}
\rho_j T_j\tQ_{ij} =  \rho_i T_iQ_{ji} \; ,
\end{equation}
and the factors $\rho_i$ are first potentials satisfying
equations
\begin{equation} \label{eq:rho-constr}
\frac{T_j\rho_i}{\rho_i} = 1 - (T_iQ_{ji})(T_jQ_{ij}) \; , i\ne j \; .
\end{equation}
\end{Prop}
\begin{proof}
Using equations
\eqref{eq:def-rho}, \eqref{eq:lin-bX} and \eqref{eq:lin-X} we obtain
\begin{equation*}
\bX_i  = - \rho_i T_i\tbX_i = 
 \frac{\rho_i}{T_j\rho_i} \left(1 - 
(T_i\tQ_{ji})(T_j\tQ_{ij})\right)
\left( \bX_i + (T_j Q_{ij})\bX_j \right)  -
\frac{\rho_i}{\rho_j}  (T_i\tQ_{ji})\bX_j \; ,
\end{equation*}
which, by comparing coefficients in front of the vectors $\bX_i$, $\bX_j$, 
leads to
equations (\ref{eq:Q-Qt}) and (\ref{eq:rho-constr}).
\end{proof}

\begin{Rem}
Since $Q_{ij}$ and $\tQ_{ij}$ are both solutions of the MQL
equations~(\ref{eq:MQL-Q}), then 
equations~(\ref{eq:def-rho})-(\ref{eq:rho-constr}) describe a special 
symmetry transformation
of equations~(\ref{eq:MQL-Q}), first found in~\cite{KoSchief2} without any
associated geometric meaning.
\end{Rem}

The RHS of equation~(\ref{eq:rho-constr}) is symmetric with
respect to the interchange of $i$ and $j$, which implies the
existence of a potential $\tau:\ZZ^N\to \RR$, such that
\begin{equation}
\rho_i = \frac{T_i\tau}{\tau} \; \; ;
\end{equation}
therefore equation~(\ref{eq:rho-constr}) defines the second potential $\tau$:
\begin{equation} \label{eq:tau}
\frac{(T_i T_j\tau)\tau}{(T_i \tau)(T_j\tau)} = 1 - (T_iQ_{ji})(T_jQ_{ij})
\; , \quad i\ne j \; .
\end{equation}
The potential $\tau$ connecting the forward and backward data:
\begin{eqnarray}
T_j(\tau\tQ_{ij}) & = & T_i(\tau Q_{ji}) \; ,
\label{eq:Q-Qt-tau}\\  T_i(\tau\tbX_i) & = & \tau\bX_i \; , \\
\tau \tH_i & = & T_i(\tau H_i) \; ,
\end{eqnarray}
is the famous $\tau$-{\it function} of the quadrilateral lattice.

\begin{Cor} The $\tau$-function representation of the MQL equations.\\
Define $\tau_{ij}$ by
\begin{equation}
\tau_{ij} = \tau Q_{ij} \; ,
\end{equation}
then equation~(\ref{eq:rho-constr}) can be rewritten as
\begin{equation} \label{eq:Hir-ij}
(T_i T_j \tau ) \tau = (T_i\tau) T_j\tau - (T_i\tau_{ji}) T_j\tau_{ij} \; ,
\end{equation}
and the MQL equations (\ref{eq:MQL-Q}) take the form
\begin{equation} \label{eq:Hir-ijk}
(T_k \tau_{ij}) \tau  = (T_k\tau) \tau_{ij} + (T_k\tau_{ik}) \tau_{kj} \; .
\end{equation}
\end{Cor}
\begin{Rem}
The $\tau$-function representation of the MQL equations was
found in~\cite{DMMMS} using the Miwa
transformation of the $\tau$-function representation of the Darboux 
equations.
\end{Rem}

We notice that, for a given lattice $\vbx$,
the forward data $\{ \bX_i, Q_{ij} \}$ are defined up to rescaling
by functions $a_i(n_i)$
\begin{equation} \label{eq:scaling-f}
\bX_i \longrightarrow a_i\bX_i \; , \quad
T_iH_i \longrightarrow \frac{1}{a_i} T_iH_i \; , \quad
T_jQ_{ij} \longrightarrow \frac{a_i}{a_j}T_jQ_{ij} \; ,
\end{equation}
expressing the freedom in the definition of the vectors $\bX_i^{(0)}$ on the
initial curves. An analogous freedom exists for the backward data
\begin{equation} \label{eq:scaling-b}
T_i\tbX_i \longrightarrow \frac{1}{b_i} T_i\tbX_i \; , \quad
\tH_i  \longrightarrow b_i \tH_i \; , \quad
T_j\tQ_{ij} \longrightarrow \frac{b_i}{b_j}T_j\tQ_{ij} \; .
\end{equation}
The corresponding rescaling of $\rho_i$ and $\tau$ is therefore
given by
\begin{equation} \label{eq:scaling-b2}
\rho_i \longrightarrow a_i b_i \rho_i \; , \quad
\tau \longrightarrow \tau \prod_{i=1}^N c_i(n_i) \; ,
\end{equation}
where
\begin{equation}
\frac{T_i c_i}{c_i}= a_i b_i \; .
\end{equation}

\begin{figure}
\begin{center}
\epsffile{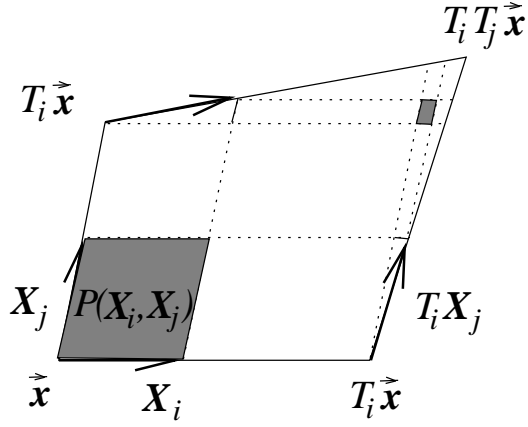}
\end{center}
\caption{Areas of two parallelograms}
\label{fig:area}
\end{figure}

Finally, we remark that the product
$(T_iQ_{ji})(T_jQ_{ij})$, which appears in the definition of the
$\tau$-function, is the ratio of
the areas of the two affine parallelograms $P(\D_i\bX_j ,
\D_j\bX_i)$ and $P(\bX_i , \bX_j)$ (see Figure~\ref{fig:area}). 

Unlike the definitions of the forward and backward rotation coefficients,
this product is
invariant with respect to their possible redefinitions
given by equation~(\ref{eq:scaling-f}). It can be seen expressing the
product, using equations (\ref{def:A-H}), in terms of the data
$A_{ij}$ as follows
\begin{equation} \label{eq:area-A}
(T_iQ_{ji})(T_jQ_{ij}) =\frac{(T_iA_{ij})( T_jA_{ji})}{(T_iA_{ij} + 1)
(T_jA_{ji} + 1)} \; .
\end{equation}

Observe finally that equation (\ref{eq:Q-Qt}) leads to
\begin{equation} \label{eq:QQ-QQ}
(T_iQ_{ji})(T_jQ_{ij}) = (T_i\tQ_{ji})(T_j\tQ_{ij}) \; ,
\end{equation}
which implies that the discussed product quantity is also the ratio of 
the areas of the backward parallelograms
$P(\tD_i T_i T_j \tbX_j , \tD_j T_i T_j\tbX_i)$ and
$P(T_i T_j\tbX_j , T_i T_j\tbX_i)$.

\section{Hyperplane lattices}
\label{sec:h-lat}

Consider a lattice $\by^*:\ZZ^N \ra (\PP^M)^*$, $N\leq M$, in the space of
hyperplanes of $\PP^M$, which we call the {\it hyperplane lattice}.
The space $(\PP^M)^*$,
called also dual space to $\PP^M$, has a natural projective structure and
{\it a priori} one expects that the
algebraic description of the quadrilateral lattices in the dual
space be the same like that of quadrilateral point lattices and,
therefore, the considerations of the previous Sections
can be applied to
hyperplane lattices without essential modifications. However,
this Section is devoted to investigate hyperplane lattices from a geometric
point of view and to make clear the geometric content of their algebraic
description.

\subsection{Quadrilateral hyperplane lattices}

The basic property of quadrilateral lattices, i.e., the planarity
of their elementary quadrilaterals, when applied to hyperplane lattices, 
can be formulated as follows.

\begin{Def} \label{def:q-h-lat}
The hyperplane lattice $\by^*:\ZZ^N\to (\PP^M)^*$ is quadrilateral if,
for any $i,j=1,\dots,N$,
$i\ne j$, the hyperplane $T_iT_j y^*$ contains the subspace
$ y^* \cap T_iy^* \cap T_jy^* $.
\end{Def}

To explain this definition notice that, if the
hyperplane lattice is given in homogeneous coordinates
by the function $\by^*:\ZZ^N\rightarrow
(\RR^{M+1})^*\setminus\{ \bzero^* \}$, then Definition~\ref{def:q-h-lat}
states that the four co-vectors $T_iT_j\by^*$, $T_i\by^*$, $T_j\by^*$,
 and $\by^*$
are linearly dependent. If $T_i\by^*$, $T_j\by^*$, $\by^*$ are linearly
independent,
then the co-vector $T_iT_j\by^*$ representing
the hyperplane $T_iT_j y^*$ is a linear combination of the co-vectors
$\by^*$, $T_i\by^*$ and $T_j\by^*$
\begin{equation} \label{eq:h-lin}
T_iT_j\by^* = \alpha T_i\by^* + \beta T_j\by^* + \gamma\by^* \; .
\end{equation}
This equation can be transformed into the Laplace equation
\begin{equation} \label{eq:h-Lapl-gen}
\D_i\D_j\by^* = (T_iA_{ij}^*)\D_i\by^* + (T_jA_{ji}^*)\D_j\by^* +
C^*_{(ij)}\by^*
\; , \quad i\ne j \;.
\end{equation}

In the affine gauge, the coefficients $\alpha$, $\beta$ and $\gamma$
of the decomposition~(\ref{eq:h-lin}) are subjected
to the constraint
\begin{equation} \label{eq:aff-con}
\alpha + \beta + \gamma = 1 \; 
\end{equation}
and the equation~(\ref{eq:h-Lapl-gen}) reduces to
\begin{equation} \label{eq:eq:h-lat-Lapl-aff}
\D_i\D_j\vby^* = (T_iA_{ij}^*)\D_i\vby^* + (T_jA_{ji}^*)\D_j\vby^*
\; , \quad i\ne j \;.
\end{equation}

\begin{Rem}
In our considerations we always assume we deal with {\it generic}
lattices, i.e., that the hyperplane $y^*$ and its forward
neighbors $T_1y^*$,\dots, $T_Ny^*$ (and backward neighbors
$T_1^{-1}y^*$,\dots, $T_N^{-1}y^*$) are in general position, i.e.,
their equations are linearly independent. In consequence, the
intersection $y^*\cap T_1y^* \cap \dots \cap T_Ny^*$ (and $y^*\cap
T_1^{-1}y^* \cap \dots \cap T_N^{-1}y^*$) is a linear subspace of
$\PP^M$ of co-dimension $N$ (of dimension $M-N$).
\end{Rem}

\begin{Ex}
Given a two dimensional quadrilateral lattice $x$ in the three dimensional
projective space, define the lattice $y^*$ of the hyperplanes passing
through $x$, $T_1 x$ and $T_2 x$. Because of the planarity of the elementary
quadrilaterals of $x$, it is easy to see that the four hyperplanes
$y^*$, $T_1 y^*$, $T_2 y^*$ and  $T_1 T_2 y^*$ intersect in the point
$T_1 T_2 x$. Therefore, the (hyper)plane lattice $y^*$ is quadrilateral.
\end{Ex}
\begin{Ex}
Correlations map quadrilateral point lattices into
quadrilateral hyperplane lattices.
\end{Ex}

\subsection{Dual systems}

We first recall that a quadrilateral lattice $\vbx'$ is called parallel to the
quadrilateral lattice $\vbx$~\cite{TQL}
(or obtained from $\vbx$ via the Combescure transformation), if the tangents
to both lattices are parallel in the corresponding points:
$\D_i\vbx' \sim \D_i\vbx$.
In consequence, the scaled tangent vectors $\bX_i'$ of the lattice $\vbx'$ 
can be choosen to be equal to those of the lattice $\vbx$: $\bX_i' = \bX_i$;
then the rotation coefficients of both lattices coincide as well: $Q_{ij} =
Q'_{ij}$, and the Lam\'e coefficients $H_i$ and $H_i'$ are solutions of the
same equation.

In this Section we will learn how to
construct quadrilateral hyperplane lattices 
using systems of parallel quadrilateral
point lattices. 

\begin{Def}
Consider a system of $M$ parallel point lattices in $\AAf^M$ 
$\vbx_{(k)}$, $k=1,\dots,M$,
whose corresponding vectors are linearly independent.
Denote by $\vby^*_{(k)}$, $k=1,\dots,M$, the system of
hyperplane lattices uniquely defined by the properties
that $\vby^*_{(k)}$ passes 
through $\vbx_{(k)}$ and is spanned
by the vectors $\vbx_{(l)}$, $l\ne k$; i.e.,
\begin{equation}
\langle \vby^*_{(k)} | \vbx_{(l)} \rangle = \delta_{kl} \; .
\end{equation}
We call such a system of hyperplane lattices the {\it dual
system} to the system of parallel point lattices $\vbx_{(k)}$.
\end{Def}

The aim of this Section is to prove that the hyperplane lattices
$\vby^*_{(k)}$ are quadrilateral hyperplane lattices.

\begin{Def} \label{def:d-Om}
Fix a basis $\{ \vbe_k \}_{k=1}^{M}$ in the ambient space $\RR^M$
and arrange the parallel system of point lattices in the
{\em matrix $\bOm$ of the system}:
\begin{equation}
\bOm = (\vbx_{(1)},\dots ,\vbx_{(M)} ) \; ;
\end{equation}
equivalently, the matrix $\bOm$ represents a linear operator
\begin{equation*}
\bOm = \sum_{k=1}^{M} \vbx_{(k)}\otimes \vbe_k^* \; ,
\end{equation*}
where $\{ \vbe_k^* \}_{k=1}^{M}$ is the dual basis of 
$\{ \vbe_k \}_{k=1}^{M}$, i.e., $\langle \vbe_k^* | \vbe_l \rangle =
\delta_{kl}$.
\end{Def}
\begin{Cor}
The components of the dual system
$\vby^*_{(k)}$ in the basis $\{ \vbe_k^* \}_{k=1}^{M}$ are given
by the rows of the matrix $\bOm^{-1}$.
\end{Cor}
Let us arrange the coefficients $H_{i(k)}$, $i=1,\dots , N$, 
$k=1,\dots , M$,
into the row-vectors
\begin{equation*}
\bX^*_i = (H_{i(1)},\dots , H_{i(M)}) \; , \quad
\bX_i^*=\sum_{k=1}^{M}H_{i(k)}\vbe_k^* ;
\end{equation*}
then 
$\bX^*_i$, $i=1,\dots,N$, form a (co)vector valued solution of the adjoint
linear problem \eqref{eq:lin-H} and
the matrix $\bOm$ can be found from equations
\begin{equation} \label{eq:D-Om}
\D_i\bOm =  \bX_i  \otimes (T_i \bX^*_i) \; .
\end{equation}
It was shown in~\cite{MDS} that the matrix $\bOm$ plays a relevant role 
in the
theory of transformations of quadrilateral lattices.

The following  Theorem, which contains, as particular cases, all the
classical transformations of a quadrilateral lattice~\cite{TQL}, 
was proven in~\cite{MDS}.
\begin{Th} \label{th:inverse-sys}
Let $Q_{ij}$, $\bX_i$, $\bX_i^*$ and $\bOm$ be defined as above; then
the following functions
\begin{equation}
Q_{ij}^{\prime} = Q_{ij} - \langle \bX_j^*|\bOm^{-1} | \bX_i \rangle
\end{equation}
solve the MQL equations, the vectors $\bX_i^\prime = \bOm^{-1}
\bX_i$, $\bX_i^{*\prime} =   \bX_i^* \bOm^{-1}$ are solutions of
the linear systems (\ref{eq:lin-X}), (\ref{eq:lin-H}) for
$Q_{ij}^\prime$, and the corresponding potential
\begin{equation}
\D_i\bOm^\prime =  \bX_i^\prime  \otimes (T_i \bX^{*\prime}_i) ,
\end{equation}
is given by
\begin{equation}
\bOm^\prime = C - \bOm^{-1} ,
\end{equation}
where $C$ is a constant operator.
\end{Th}

Denote by $\vbx^*_{(k)}$ the rows of $\bOm$;
then
\begin{equation*}
\bOm = \sum_{k=1}^{M} \vbe_k \otimes \vbx^*_{(k)} \; .
\end{equation*}

\begin{Lem}
The rows $\vbx^*_{(k)}$ of the matrix $\bOm$ represent
a system of co-parallel quadrilateral hyperplane lattices, which
we call the adjoint system to $\vbx_{k}$.
\end{Lem}
\begin{proof}
Let us rewrite equation (\ref{eq:D-Om})
in a backward form
\begin{equation} \label{eq:D-Om-b}
\tilde{\D}_i\bOm = \bX^*_i \otimes (T_i^{-1} \bX_i) \; ,
\end{equation}
which gives
\begin{equation}
\tD_i \vbx^*_{(k)} = (T_i^{-1}H^*_{i(k)})\bX^*_i \; ,
\end{equation}
where $H^*_{i(k)}$ is the $k$-th component of the vector $\bX_i$.
Comparing equations (\ref{eq:b-H-X}), (\ref{eq:lin-bH}) and 
(\ref{eq:lin-X})
we infer that the co-vectors $\vbx^*_{(k)}$ satisfy
the backward Laplace equations
\begin{equation} \label{eq:x*-Lapl-1}
\tilde{\D}_i\tilde{\D}_j\vbx^*_{(k)} = (T_i^{-1}\tilde{A}_{ij(k)}^*)
\tilde{\D}_i\vbx^*_{(k)} +
(T_j^{-1}\tilde{A}_{ji(k)}^*)\tilde{\D}_j\vbx^*_{(k)}
\; , \quad i\ne j \; ,
\end{equation}
where
\begin{equation*}
\tilde{A}_{ij(k)}^*= \frac{\tilde{\D}_j H^*_{i(k)}}{H^*_{i(k)}} \; ,
\end{equation*}
and therefore (see Proposition~\ref{prop:b-f}) also the forward Laplace 
equations.

Finally, since $\tD_i\vbx^*_{(k)} \sim \tD_i\vbx^*_{(l)}$, then the 
corresponding
co-dimension $2$ subspaces $x^*_{(k)}\cap T_i^{-1} x^*_{(k)}$ and
$x^*_{(l)}\cap T_i^{-1} x^*_{(l)}$ of hyperplane lattices are co-parallel
in the sense of Definition~\ref{def:co-parallelity}.
\end{proof}

\begin{Rem}
Given the parallel system $\vbx_{(k)}$, $k=1,\dots , M$, the
corresponding adjoint system $\vbx^*_{(k)}$ 
is given up to a fixed basis used to define $\bOm$; on the
contrary, the dual system of hyperplane lattices $\vby_{(k)}^*$ is
given uniquely.
\end{Rem}

\begin{Cor}
Notice that the forward rotation coefficients  of 
the system $\vbx_{(k)}$
are the backward rotation coefficients of the system $\vbx^*_{(k)}$:
$Q_{ij}= \tQ_{ij}^*$.
\end{Cor}

Combining the above Lemma with Theorem~\ref{th:inverse-sys} we get
the following

\begin{Th}
The hyperplane lattices $\vby^*_{(k)}$ of the dual system  to the
system of parallel
quadrilateral point lattices $\vbx_{(k)}$ are co-parallel quadrilateral
hyperplane lattices.
\end{Th}

\subsection{The adjoint and conjugate lattices}

\begin{Def} \label{def:adj-h-p-latt}
The quadrilateral point lattice $\vbx:\ZZ^N\to\AAf^M$ and the 
quadrilateral hyperplane lattice
$\vbx^*:\ZZ^N\to(\AAf^M)^*$ are called adjoint if the forward rotation
coefficients of the point lattice are backward rotation coefficients
of the hyperplane lattice.
\end{Def}
\begin{Cor}
Equivalently, the forward rotation
coefficients of the hyperplane lattice are backward rotation coefficients
of its adjoint point lattice.
\end{Cor}

\begin{Def} \label{def:conj-h-p-latt}
The point lattice $x:\ZZ^N\to\PP^M$ and the hyperplane lattice
$y^*:\ZZ^N\to(\PP^M)^*$ are called conjugate if there exists a one to
one correspondence between both lattices such that
the points $x$ of the point lattice belong to the corresponding
hyperplanes $y^*$ of the hyperplane lattice.
\end{Def}

\begin{Cor}
Observe that this notion is self-dual in the sense of the standard
duality between $\PP^M=((\PP^M)^*)^*$ and $(\PP^M)^*$.
\end{Cor}

\begin{Rem} We are interested only
in a situation in which $x$ is a quadrilateral point lattice and
$y^*$ is a quadrilateral hyperplane lattice.
\end{Rem}

The notion of conjugacy between point lattices
and hyperplane lattices is the natural generalization of the notion of
conjugacy between point lattices and rectilinear congruences
(line lattices with any two neighbouring lines coplanar)
introduced in~\cite{TQL}.
As it was also shown in ~\cite{TQL}, the lattices parallel to $\vbx$ 
describe transversal congruences
conjugate to the quadrilateral point lattice $x$.  Moreover, 
the tangent congruences can be also obtained in this way
via singular limits.
\begin{Cor}
Given a quadrilateral point lattice $x$ in $\PP^M$ and given $(M-1)$
linearly independent congruences conjugate to $x$, then the hyperplane
lattice $y^*$ conjugate to $x$ and spanned by the lines of these 
congruences
is quadrilateral.
\end{Cor}

\subsection{The complementary lattice}

The linear system \eqref{eq:lin-X} describes the variation of the normalized 
tangent vectors $\bX_i$ of a quadrilateral point lattice in directions 
$j\ne i$, and leads to the MQL equations~\eqref{eq:MQL-Q}. In this Section
we study the variation of the vectors $\bX_i$ in the corresponding $i$-th 
directions of the lattice.
Discussion of such variations naturally leads to the definition of
a hyperplane lattice, which will be called the {\it complementary} lattice.

Consider the quadrilateral lattice $\vbx:\ZZ^N\rightarrow\RR^M$ with the
given set of 
tangent vectors $\bX_i$, $i=1,\dots,N$, and the corresponding set of the
Lam\'e and rotation coefficients $H_i$, $Q_{ij}$, $i,j=1,\dots,N$ 
satisfying equations \eqref{def:HX}--\eqref{eq:lin-H}. Let us find
$M-N$ new solutions $Q_{ai}$, $a=N+1,\dots,M$, $i=1,\dots,N$ of the 
adjoint linear system~\eqref{eq:lin-H}, i.e.,
\begin{equation} \label{eq:Qai}
\D_i Q_{aj} = (T_i Q_{ai})Q_{ij}, 
\end{equation}
and let us define $M-N$ vectors $\bX_a$, $a=N+1,\dots,M$, via an analogue 
of equations~\eqref{def:HX}
\begin{equation} \label{eq:Xa}
\D_i\bX_a = (T_i Q_{ai}) \bX_i .
\end{equation}
\begin{Rem}
The vectors $\bX_a$, $a=N+1,\dots,M$, are the Combescure transforms of the
lattice $\vbx$, but it was not accidental that we gave to
equations~\eqref{eq:Qai} and~\eqref{eq:Xa}
the form of the Darboux equations~\eqref{eq:MQL-Q} and of the linear
problem~\eqref{eq:lin-X}.
\end{Rem}
When the full set of vectors $\bX_k$, $k=1,\dots,M$, is linearly 
independent,
we obtain, in each point of the lattice $\vbx$, a basis of the whole 
space
$\RR^M$; this type of basis along a quadrilateral lattice has been
considered already in~\cite{DMS} and can be called the {\it
extended basis} along the lattice. By ${\tbY}_k^*$, $k=1,\dots,M$, we 
denote 
the dual basis in $(\RR^M)^*$
\begin{equation} \label{eq:tYkXl}
\langle {\tbY}_k^* | \bX_\ell \rangle = \delta_{k\ell}, \qquad
k,\ell=1,\dots, M.
\end{equation}
The linear system~\eqref{eq:lin-X} describes the decomposition of $T_i \bX_j$, 
$i\ne j$; let us decompose $T_i\bX_i$ in the full basis
\begin{equation} \label{eq:DiXi}
\D_i\bX_i = \tP^*_i\bX_i -\sum_{k\ne i,k=1}^M \tP^*_{ik}\bX_k .
\end{equation}
We will study properties of the coefficients $\tP^*_i$, $\tP^*_{ij}$ and their 
relation
to previously introduced objects.
\begin{Prop}
The vectors $\tbY^*_k$ satisfy equations
\begin{equation} \label{eq:DitYk}
\D_i\tbY^*_k = (T_i\tbY^*_i)\tP^*_{ik}, \quad i\ne k,
\quad i=1, \dots, N, \quad k=1, \dots , M,
\end{equation}
\begin{equation} \label{eq:DitYi}
\D_i\tbY^*_i = - (T_i\tbY^*_i)\tP^*_i - \sum_{k\ne i, k=1}^{M} 
(T_i\tbY^*_k)(T_i
Q_{ki}), \quad i=1, \dots, N .
\end{equation}
\end{Prop}
\begin{proof}
Assume a decomposition of $\D_i\tbY^*_\ell$ in the basis $T_i\tbY^*_\ell$,
$\ell=1,\dots ,M$
\begin{equation}
\D_i\tbY^*_\ell = \sum_{k=1}^M \Gamma_{i\ell}^k (T_i\tbY^*_k),
\end{equation}
where
\begin{equation*}
\Gamma_{i\ell}^k = \langle \D_i\tbY^*_\ell | T_i\bX_k \rangle ,
\quad i=1, \dots, N, \quad k,\ell=1, \dots , M.
\end{equation*}
Using equation~\eqref{eq:tYkXl} we obtain that
\begin{equation}
\Gamma_{i\ell}^k = - \langle \tbY^*_\ell | \D_i\bX_k \rangle, 
\end{equation}
which, together with equations~\eqref{eq:tYkXl}, 
\eqref{eq:DitYk} and~\eqref{eq:DiXi}
and~\eqref{eq:lin-X}, concludes the proof.
\end{proof}
\begin{Cor} \label{Cor:supp1}
Equation~\eqref{eq:DitYk} can be splitted into the standard
backward linear problem 
\begin{equation} \label{eq:DitYj}
\D_i\tbY^*_j = (T_i\tbY^*_i)\tP^*_{ij}, \quad i\ne j,
\quad i=1, \dots, N,
\end{equation}
and the backward linear equations for the supplementary covectors 
\begin{equation} \label{eq:DitYa}
\D_i\tbY^*_a = (T_i\tbY^*_i)\tP^*_{ia},
\quad i=1, \dots, N, \quad a=N+1, \dots , M.
\end{equation}
The compatibility
condition of these equations gives the Darboux equations for the backward
rotation coefficients $\tP^*_{ij}$, $i\ne j=1,\dots ,N$ 
\begin{equation} \label{eq:MQL-P}
\D_k\tP^*_{ij} = (T_k\tP^*_{ik})\tP^*_{kj}, \quad k\ne i,j=1, \dots N,
\end{equation}
and the supplementary backward linear equations
\begin{equation} \label{eq:P}
\D_i\tP^*_{ja} = (T_i\tP^*_{ij})\tP^*_{ia}, \quad i\ne j=1, \dots N, 
\quad a=N+1,
\dots ,M.
\end{equation}
\end{Cor}
\begin{Cor} \label{Cor:supp2}
The compatibility of equations~\eqref{eq:DitYk} and~\eqref{eq:DitYi} gives
\begin{equation} \label{eq:Pi}
\tP^*_i = T_iQ_{ii} - \tP^*_{ii}, \quad i=1,\dots ,N,
\end{equation}
where $Q_{ii}$ (and similarly $\tP^*_{ii}$)
are potentials defined in~\cite{DMS} for any solution of the MQL system,
by the
equations
\begin{equation} \label{eq:Qii}
\D_jQ_{ii}= (T_jQ_{ij})Q_{ji}, \quad \D_j\tP^*_{ii}= (T_j\tP^*_{ij})
\tP^*_{ji},
\quad j\ne i .
\end{equation}
Moreover, from the same compatibility, we
obtain the following equation
\begin{equation} \label{eq:PQ}
\D_iQ_{ij} + \tD_j\tP^*_{ij} - \tP^*_i Q_{ij} + \tP^*_{ij}(T_j^{-1}\tP^*_j)
+ \sum_{k\ne i,j; k=1}^M \tP^*_{ik}Q_{kj}=0, \quad i\ne j.
\end{equation}
\end{Cor}
To make the above considerations symmetric we define a hyperplane lattice
which has the vectors $\tbY^*_i$, $i=1,\dots ,N$, as normalized
backward tangent vectors, and $\tP^*_{ij}$, $i\ne j=1,\dots ,N$ as backward
rotation coefficients. 
\begin{Def}
Given the quadrilateral lattice $\vbx:\ZZ^N\rightarrow \RR^M$ together 
with its extended frame $\bX_k$ and its dual $\tbY_k$, $k=1,\dots,M$,
define the {\it complementary lattice} of $\vbx$ as 
via solution of the following
compatible equations
\begin{equation}
\D_i\vby^* = (T_i\tbY^*_i)\tF^*_i , i=1,\dots,N,
\end{equation}
where $\tF^*_i$, $i=1,\dots,N$, is a solution of the
system \eqref{eq:P}, interpreted now as the adjoint of the linear
system~\eqref{eq:DitYj}
\begin{equation}
\D_j\tF^*_i = (T_j\tP^*_{ji})\tF^*_j , \quad i\ne j=1,\dots,N.
\end{equation}
\end{Def}
\begin{Rem}
The additional vectors $\tbY^*_a$ and functions $\tP^*_{ia}$,
$a=N+1,\dots,M$ play a role similar to that of $\bX_a$ and $Q_{ai}$.
\end{Rem}
By simple calculation one can obtain the following result.
\begin{Prop}
The functions $v_k=\langle \vby^* | \bX_k \rangle$, $k=1,\dots M$
satisfy equations
\begin{align} \label{eq:lin-X-ex}
\D_iv_k &= (T_iQ_{ki})v_i, \qquad k\ne i \\
\label{eq:Divi-F}
\D_iv_i &= \tF^*_i + \tP^*_i v_i - \sum_{k\ne i}\tP^*_{ik}v_k .
\end{align}
Similary, functions $\tv^*_k=\langle \tbY^*_k | \bx \rangle$, $k=1,\dots M$
satisfy equations
\begin{align}
\D_i\tv^*_k &= (T_i\tv^*_i)\tP_{ik}, \qquad k\ne i \\
\D_i\tv^*_i &= (T_iH_i)  - \tP^*_i (T_i\bv^*_i) - \sum_{k\ne i}(T_iQ_{ki})
\bv^*_k .
\end{align}
\end{Prop}
Finally, we present a Theorem, which can be proved by simple algebra using
formulas of Corollaries~ \ref{Cor:supp1} and~\ref{Cor:supp2}, and which
contains a geometric characterization of the complementary lattice.
\begin{Th}
Consider the quadrilateral lattice $\vbx$ with the extended frame $\bX_k$,
$k=1,\dots,M$, and consider a scalar solution $v_k$ of the extended linear
system~\eqref{eq:lin-X-ex}. The hyperplane lattice $\vby^* =
\sum_{k=1}^M v_k\tbY^*_k$, whose hyperplanes pass through the $M$ points  
$\frac{1}{v_k}\bX_k$, is a complementary lattice of $\vbx$. Its backward
Lam\'e coefficients $\tF^*_i$, $i=1,\dots,N$ can be obtained via formulas
\eqref{eq:Divi-F}. 
\end{Th}

\begin{Rem}
In the continuous limit, for $N=M=3$, and with the identification of planes in
$\EE^3$ as points (via polarity), our complementary hyperplane lattices
reduce to the "syst\`emes compl\'ementaires d'un syst\`eme conjugu\'e"
considered by Darboux~\cite[Chapt. III]{DarbouxOS}. 
\end{Rem}

\section{The Symmetric Lattice}

\label{sec:symm-latt}

\begin{Def} A quadrilateral lattice $\vbx$ is {\em symmetric}
iff its forward rotation coefficients are its backward rotation
coefficients as well; i.e.
\begin{equation} \label{eq:Q=tQ}
\tQ_{ij} = Q_{ij} \; .
\end{equation}
\end{Def}

The considerations of Section~\ref{sec:backward} imply the following
characterization.

\begin{Prop} A quadrilateral lattice is symmetric iff,
for a given set of rotation coefficients $Q_{ij}$,
there exists a $\tau$--function of the lattice such that
\begin{equation} \label{eq:symm}
T_i(\tau Q_{ji})=T_j(\tau Q_{ij}) \; , \quad i\ne j \; ,
\end{equation}
or equivalently, in terms of the corresponding first potentials $\rho_i$,
\begin{equation} \label{eq:symm-rho}
\rho_iT_iQ_{ji}=\rho_jT_jQ_{ij} .
\end{equation}
\end{Prop}

\begin{Rem}
Due to equations \eqref{eq:scaling-f}-\eqref{eq:scaling-b2}, the above 
definition
is independent of the particular choice of the rotation coefficients
$Q_{ij}$.
\end{Rem}

It turns out that
\begin{Prop}
The symmetric lattice is an integrable reduction of the quadrilateral
lattice.
\end{Prop}
\begin{proof}
Recall that, from a geometric point of view, the integrability
of a reduction means that, if the reduction condition is satisfied on
the initial
surfaces, then it must propagate in the construction of the lattice.

As it was shown in \cite{MQL} the solution $Q_{ij}$
of the MQL equations ~(\ref{eq:MQL-Q}) is fixed
by the values of the rotation coefficients $Q_{ij}^{(0)}$ on the initial
surfaces. Therefore, if  $Q_{ij}^{(0)}=\tQ_{ij}^{(0)}$ on the 
initial surfaces, then they are equal $Q_{ij}=\tQ_{ij}$ in the whole
lattice, since the backward rotation coefficients $\tQ_{ij}$ satisfy the
same equations as $Q_{ij}$.

The algebraic content of this result is instead expressed by the
following equation
\begin{equation} \label{eq:symmconstr}
T_kC_{ij}^S=C_{ij}^S+(T_kQ_{jk})C_{ik}^S-(T_kQ_{ik})C_{jk}^S \; ,\quad
i\ne j\ne k,
\end{equation}
where
\begin{equation} \label{def:C^S}
C_{ij}^S:=\rho_iT_iQ_{ji}-\rho_jT_jQ_{ij} \; ;
\end{equation}
equation (\ref{eq:symmconstr}) is a simple consequence of the MQL
equations~(\ref{eq:MQL-Q}) and of equation~(\ref{eq:rho-constr}).
Again we see that, if the constraint (\ref{eq:symm}) is satisfied
on the initial surfaces (the RHS of equation (\ref{eq:symmconstr})
is zero), then it propagates transversally through the whole
lattice (the LHS of equation (\ref{eq:symmconstr}) is zero).
\end{proof}

There exists an interesting geometric characterization of the
symmetric lattice, which follows from the interpretation of the
condition $\tQ_{ij}=Q_{ij}$.

\begin{figure}
\begin{center}
\epsffile{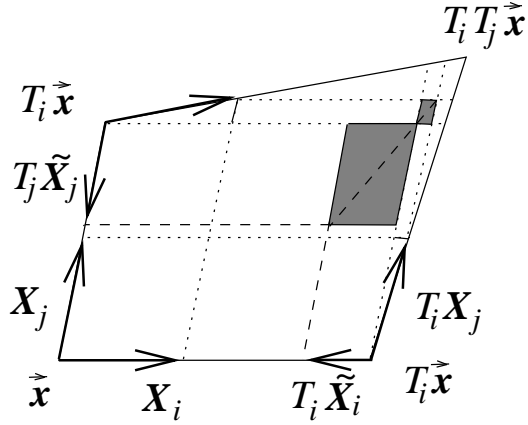}
\end{center}
\caption{Similarity of two parallelograms}
\label{fig:symm}
\end{figure}

\begin{Lem}
The forward and backward rotation coefficients describing an
elementary quadrilateral $\{ \vbx, T_i\vbx, T_j\vbx, T_iT_j\vbx \}$ 
are equal if and only if the parallelograms $P(T_i\tbX_i , T_j\tbX_j)$
and $P(\D_i\bX_j , \D_j\bX_i)$ of the quadrilateral are similar.
\end{Lem}
\begin{proof}
The quadrilateral with the initial vertex is described by the
following rotation coefficients: $T_iQ_{ji}$, $T_jQ_{ij}$,
$T_i\tQ_{ji}$ and $T_j\tQ_{ij}$ connected by equation
\eqref{eq:Q-Qt}. Since
\begin{equation}
\D_i\bX_j=-(T_iQ_{ji})\rho_i T_i\tbX_i \; ,
\end{equation}
then the parallelograms $P(T_i\tbX_i , T_j\tbX_j)$ and
$P(\D_i\bX_j , \D_j\bX_i)$ are similar (see Figure~\ref{fig:symm}) if and only if
\begin{equation}
\rho_j(T_jQ_{ij}) = \rho_i(T_iQ_{ji}) \; ,
\end{equation}
which means, due to~(\ref{eq:Q-Qt}), that the backward and forward
$Q$'s are equal.
\end{proof}

\begin{Prop}
A quadrilateral lattice is symmetric iff, 
for a given set of the forward tangent vectors $\bX_i$ of the lattice,
there exists a complementary set of the backward tangent vectors $\tbX_i$ 
such that the parallelograms
$P(T_i\tbX_i , T_j\tbX_j)$ and $P(\D_i\bX_j , \D_j\bX_i)$ are
similar.
\end{Prop}
\begin{Rem}
Due to equations \eqref{eq:scaling-f}-\eqref{eq:scaling-b2} the above
characterization of the symmetric lattice
is independent of a particular choice of the vectors $\bX_i$.
\end{Rem}

Integrability of the symmetric lattice can be formulated as follows
\begin{Cor}
If the system of initial quadrilateral surfaces admits a compatible set
of forward--backward data such that $P(T_i\tbX_i , T_j\tbX_j)$ and
$P(\D_i\bX_j , \D_j\bX_i)$ are similar, then the similarity of the
parallelograms holds in the whole quadrilateral lattice.
\end{Cor}

\begin{Rem}
Notice that, in order to define the symmetric lattice, we need to know what 
similar parallelograms
are.
\end{Rem}

The solution of the MQL equations for an $N$ dimensional symmetric lattice
depends on $\binom{N}{2}$ arbitrary functions of two variables, i.e., one
half of the arbitrary functions parametrizing the solution of the MQL
equations for generic $N$ dimensional quadrilateral lattice (see
Section~\ref{sec:for-ql}).
Given a symmetric lattice equipped with a compatible set of forward and
backward data, denote the similarity factor between the parallelograms 
$P(T_i\tbX_i , T_j\tbX_j)$ and $P(\D_i\bX_j , \D_j\bX_i)$ by
$\sigma_{(ij)} =\sigma_{(ji)}$
\begin{equation}
\D_i\bX_j  =  \sigma_{(ij)}T_i\tbX_i \; , \quad
\D_j\bX_i  =  \sigma_{(ij)}T_j\tbX_j \; , \quad i\ne j \; ,
\end{equation}
then
\begin{equation}
\sigma_{(ij)} = - \rho_i T_iQ_{ji} = - \rho_j T_j Q_{ij}   \; .
\end{equation}
Therefore, to construct the initial $(i,j)$-surface of a symmetric
lattice, one gives two arbitrary intersecting $i$- and $j$- curves
and, on them, the tangent vectors $\bX_i^{(0)}$, $\bX_j^{(0)}$ and the
factors  $\rho_i^{(0)}$, $\rho_j^{(0)}$; one finally gives
$\sigma_{(ij)}=\sigma_{(ji)}$ as functions of $(n_i,n_j)$.

The descriptions of the symmetric lattice presented above are not explicit.
Indeed they involve statements about the existence of suitable potentials.
There exists however another characterization of the symmetric lattice
in terms of the forward rotation coefficients only.
\begin{Th}
A quadrilateral lattice is symmetric iff, for different indices
$i,j,k$, its rotation
coefficients satisfy the following constraint
\begin{equation}\label{eq:symm-cons-Q}
(T_iQ_{ji})(T_jQ_{kj})(T_kQ_{ik})=(T_jQ_{ij})(T_iQ_{ki})(T_kQ_{jk}).
\end{equation}
\end{Th}
In the proof we will use two simple facts (equations
\eqref{eq:def-R}-\eqref{eq:R-R} and \eqref{eq:MQL-Q-new} below) 
valid for a generic
quadrilateral lattice.

For a given set of the compatibile forward and backward rotation
coefficients $Q_{ij}$ and $\tQ_{ij}$, define functions $R_{ij}$ as
\begin{equation} \label{eq:def-R}
R_{ij} = \frac{T_jQ_{ij}}{T_j\tQ_{ij}} \; ;
\end{equation} 
then from equations \eqref{eq:Q-Qt} it follows that
\begin{equation} \label{eq:R-R}
R_{ij} = \frac{1}{R_{ji}} .
\end{equation}
The MQL equation \eqref{eq:MQL-Q} can be written as
\begin{equation} 
T_kT_jQ_{ij} = T_jQ_{ij} + (T_kT_jQ_{ik})\: (T_jQ_{kj}) ,
\end{equation}
which implies
\begin{equation}
T_kT_jQ_{ik} = \frac{T_kT_jQ_{ij} - T_jQ_{ij}}{T_jQ_{kj}} ;
\end{equation}
interchanging the indices $j$ and $k$ in the second equation and eliminating
$T_kT_jQ_{ij}$ we obtain
\begin{equation} \label{eq:MQL-Q-new}
\frac{T_k T_j Q_{ik}}{T_k Q_{ik}} = 
\frac{1 + \frac{(T_k Q_{jk}) (T_j Q_{ij})}{T_k Q_{ik}}}{ 1 - (T_jQ_{kj})\:
(T_k Q_{jk})}  .
\end{equation}
\begin{proof}[Proof of the Theorem]
The implication \eqref{eq:symm-rho} $\Rightarrow$ \eqref{eq:symm-cons-Q} 
is obvious. Let us concentrate on the opposite implication.

Let us start from any set of backward rotation coefficients $\tQ_{ij}$
related with $Q_{ij}$ via equations \eqref{eq:def-rho} \eqref{eq:Q-Qt};
the condition \eqref{eq:symm-cons-Q} implies
\begin{equation}
\frac{(T_k Q_{jk})( T_j Q_{ij})}{T_k Q_{ik}} = 
\frac{(T_k \tQ_{jk})( T_j \tQ_{ij})}{T_k \tQ_{ik}} ,
\end{equation}
which, together with \eqref{eq:MQL-Q-new} and with the corresponding
formula satisfied by the backward rotation coefficients $\tQ_{ij}$,
gives, for $j$ different from $i$ and $k$,
\begin{equation}
T_j R_{ik} = R_{ik} ;
\end{equation}
i.e., $R_{ik}$ is a function of $n_i$ and $n_k$ only.
This, together with condition \eqref{eq:symm-cons-Q} written in terms
of $R_{ij}$, as
\begin{equation}
R_{ij} R_{jk} R_{ki} = 1
\end{equation}
and with equation \eqref{eq:R-R}, implies the existence of functions $a_i(n_i)$
such that
\begin{equation}
R_{ij}(n_i,n_j) = \frac{a_i(n_i)}{a_j(n_j)} .
\end{equation}
We use the functions $a_i$ to redefine the potentials $\rho_i$ and obtain
new backward rotation coefficients $\tQ_{ij}$ satisfying
$Q_{ij} = \tQ_{ij}$.
\end{proof}

The above characterization of the symmetric lattice 
works only when the dimension of the lattice
is greater then two. In the following Proposition we present an analogous
criterion for $N=2$, which can be useful, for example,
to check directly if the initial
quadrilateral surfaces are symmetric.
\begin{Prop}
A two dimensional quadrilateral lattice is symmetric iff
the function
\begin{equation} 
r_{ij} = \frac{T_j Q_{ij}}{T_i Q_{ji}} , \qquad i\ne j ,
\end{equation}
satisfies equation
\begin{equation}\label{eq:symm-QQ}
\frac{(T_i T_j r_{ij}) r_{ij}}{(T_i r_{ij}) (T_jr_{ij})} = 
\frac{T_i (1 - T_iQ_{ji} \:T_jQ_{ij}) }{T_j (1 - T_iQ_{ji} \:T_jQ_{ij}) } .
\end{equation}
\end{Prop}
\begin{proof} 
The implication from \eqref{eq:symm-rho} to \eqref{eq:symm-QQ} is trivial. 
To prove that the condition \eqref{eq:symm-QQ} is sufficient we notice
that, in terms of $R_{ij}$,
it can be rewritten as
\begin{equation}
(T_iT_j R_{ij}) R_{ij} = (T_i R_{ij})(T_j R_{ij}) ,
\end{equation}
which leads again to
\begin{equation}
R_{ij}(n_i,n_j) = \frac{a_i(n_i)}{a_j(n_j)} .
\end{equation}
\end{proof}
\begin{Rem}
In order to check the symmetry condition for the initial surfaces we use 
the criterion \eqref{eq:symm-QQ} supplemented by \eqref{eq:symm-cons-Q} in
the points where the initial surfaces meet.
\end{Rem}

As we have anticipated, the constraints discussed in this paper allow
one to establish a connection between quadrilateral point lattices
and their duals, the quadrilateral hyperplane lattices. The following
proposition
describes this connection in the case of the symmetry constraint.

\begin{Prop} \label{prop:Om-symm}
Given a system of parallel quadrilateral
lattices $\{\vbx_{(k)}\}_{k=1}^M$ and the associated matrix $\bOm$
defined with respect to an orthonormal basis $\{ \vbe_k
\}_{k=1}^{M}$, $\vbe_k \cdot \vbe_l = \delta_{kl}$, then the
following properties are equivalent.\\ i) The matrix $\bOm$ of the
system is symmetric:
\begin{equation}\label{eq:Omegasymm}
\bOm=\bOm^T.
\end{equation}
ii) The polar hyperplane ${\cal P}({\vbx_{(k)}})$ of the point lattice
$\vbx_{(k)}$ coincides with the hyperplane lattice $\vbx^*_{(k)}$:
\begin{equation}
{\cal P}({\vbx_{(k)}})={\vbx^*_{(k)}} \; ,\quad k=1,..,M \; .
\end{equation}
iii) The lattices $\vbx_{(k)}$, $k=1,\dots,M$ are symmetric.
Furthermore the associated tangent vectors $\bX_i$ and $\bX_i^*$
are related in the following way
\begin{equation}\label{eq:symmX}
\bX_i^T = \rho_i(T_i\bX^*_i) , \quad i=1,\dots ,N.
\end{equation}
\end{Prop}

\begin{proof}
i) $\Leftrightarrow$ ii)
The equivalence of i) and ii) follows
immediately from the definitions of the potential matrix  $\Omega$
and of the polar transformation $\cal P$. \\ i) $\Rightarrow$
iii). The application of $\Delta_i$ to equation
(\ref{eq:Omegasymm}) gives the equations
\begin{equation}
\bX_i \otimes T_i\bX_i^* = T_i\bX_i^{*T} \otimes \bX_i^T ,
\end{equation}
which imply equations $\bX_i^T=\gamma_i T_i\bX_i^*$ for some
proportionality factor functions $\gamma_i$. The linear
problem~\eqref{eq:lin-X} and its adjoint~\eqref{eq:lin-H} satisfied by
$\bX_i$ and $\bX_i^*$ imply that $\gamma_i$ satisfy equations
(\ref{eq:rho-constr}) (which allows to identify $\gamma_i$ with
$\rho_i$) and lead to the symmetry condition (\ref{eq:symm}).\\
iii) $\Rightarrow$ i) Following a similar strategy, one can show
that
\begin{equation}
\D_i(\bOm-\bOm^T)=0,\;\;i=1,..,N,
\end{equation}
which implies (\ref{eq:Omegasymm}), up to some constant of integration.
\end{proof}
\begin{Cor}
A quadrilateral lattice $\vbx$ is symmetric iff it is adjoint to its own
polar.
\end{Cor}
\begin{Rem}
In the continuous limit \eqref{eq:limit-Q} the symmetric
quadrilateral lattice reduces to a {\em symmetric conjugate net},
for which the rotation coefficients $\beta_{ij}$ satisfying the
Darboux equations \eqref{eq:Darboux} are symmetric
\begin{equation}  \label{eq:symm-cont}
\beta_{ij} = \beta_{ji} \; .
\end{equation}
In fact, one should allow for the less restricitve condition
\begin{equation} \label{eq:symm-cont2}
\beta_{ij}(u) = \frac{a_i(u_i)}{a_j(u_j)}\beta_{ji}(u) \; ,
\end{equation}
which gives 
\eqref{eq:symm-cont}, after an admissible
rescaling of the data.

The continuous limit of the criterion
\eqref{eq:symm-cons-Q} 
\begin{equation}
\beta_{ij} \beta_{jk} \beta_{ki} = \beta_{ji} \beta_{kj} \beta_{ik} 
\end{equation}
is equivalent to \eqref{eq:symm-cont2}.
\end{Rem}
\section{The Circular Lattice}
\label{sec:circ-latt}

The discrete analogue of an $N$ - dimensional orthogonal system of
coordinates
is the circular lattice.

\begin{Def}
A quadrilateral lattice is circular if and only if
any elementary quadrilateral is inscribed in a
circle.
\end{Def}

An elementary characterization of circular quadrilaterals states
that, if a circular quadrilateral is convex, then the sum of its
opposite angles is $\pi$; when the quadrilateral is skew, then its
opposite angles are equal. This leads to a convenient
characterization of a circular lattice~\cite{DMS}.
\begin{Prop}
A quadrilateral lattice is circular if and only if:
\begin{equation}
\cos \angle (\bX_i , T_i\bX_j) + \cos \angle
(\bX_j , T_j\bX_i) = 0 \; 
\end{equation}
or, equivalently,
\begin{equation} \label{eq:circularity1}
\bX_i\cdot T_i\bX_j+\bX_j\cdot T_j\bX_i =0 \; , \quad i\neq j \; .
\end{equation}
\end{Prop}
It turns out that~\cite{CDS,DMS}
\begin{Prop}
The circular lattice is an integrable reduction of the quadrilateral
lattice.
\end{Prop}
\begin{proof}
The proof consists in showing that the circularity property is an
admissible constraint for the quadrilateral lattice; i.e., once
imposed on the initial surfaces, it propagates transversally
through the lattice. This was shown in~\cite{CDS} using purely
geometric means. The algebraic proof is instead based on the
following formula
\begin{equation}\label{eq:circconstr}
T_kC^{\circ}_{ij}=C^{\circ}_{ij}+(T_iT_kQ_{jk})C^{\circ}_{ik}+
(T_jT_kQ_{ik})C^{\circ}_{jk} \; ,\quad i\ne j\ne k\ne i,
\end{equation}
where
\begin{equation} \label{def:C-circularity}
C^{\circ}_{ij}:=\bX_i\cdot T_i\bX_j+\bX_j\cdot T_j\bX_i \; , \quad
i\neq j \; ,
\end{equation}
which is a direct consequence of
equations (\ref{eq:lin-X})-(\ref{eq:MQL-Q}). We see that, if the
circularity constraint (\ref{eq:circularity1}) is satisfied on the
initial surfaces (the RHS of (\ref{eq:circconstr}) is zero), then
it propagates transversally through the lattice (the LHS of
(\ref{eq:circconstr}) is zero).
\end{proof}

\begin{Cor}
1) The circularity constraint (\ref{eq:circularity1}) implies the following
formula~\cite{DMS}:
\begin{equation} \label{eq:circularity2}
{T_i|\bX_j|^2\over |\bX_j|^2}=1-(T_iQ_{ji})(T_iQ_{ji})  \; 
\end{equation}
which, compared with equations
(\ref{eq:rho-constr})-(\ref{eq:tau}), allows to fix, without loss
of generality, the backward formulation of the circular lattice in
the following way:
\begin{equation} \label{eq:circularity3}
|\bX_i|^2=  \rho_i=  {T_i\tau \over \tau} \quad \Rightarrow \quad
|T_i\tbX_i|^2 =  1/\rho_i = \frac{\tau}{T_i\tau}.
\end{equation}
2) The circularity constraint
(\ref{eq:circularity1}), written in terms of the backward data of
the lattice, reads as follows:
\begin{equation} \label{eq:circularityb}
\tilde{C}^\circ_{ij}:= \tbX_i \cdot T_i^{-1}\tbX_j + \tbX_j \cdot
T_j^{-1}\tbX_i = 0 \; .
\end{equation}
\end{Cor}
\begin{proof} 1) Equation~(\ref{eq:circularity3}) is a strightforward
consequence of equation~(\ref{eq:circularity1}) and has been found
in~\cite{DMS}.\\
2) Equation~(\ref{eq:circularityb}) follows from the equalities
\begin{eqnarray*}  \label{eq:circularitybf}
C_{ij}^\circ & = & \rho_i \rho_j   \left(
2(T_i\tbX_i)\cdot (T_j\tbX_j)+
(T_i\tQ_{ji})|T_j\tbX_j|^2+
(T_j\tQ_{ij})|T_i\tbX_i|^2 \right) = \\
 & =& \rho_i \rho_j \left( 1 - (T_iQ_{ji})(T_jQ_{ij})\right) T_iT_j
 \tilde{C}_{ij}^\circ \; .
\end{eqnarray*}
The first equality follows from rewriting $C_{ij}^\circ$ in
terms of the backward data; the second equality follows from
equations
\begin{equation}
T_iT_j\tbX_i =  \left( 1 - (T_iQ_{ji})(T_jQ_{ij})\right)^{-1}
\left( T_i\tbX_i + (T_i\tQ_{ji})T_j\tbX_j \right) \; , \quad i\neq
j \; ,
\end{equation}
straightforward consequence of~\eqref{eq:lin-bX}.
\end{proof}
Other two convenient characterizations of the circular lattice are
contained in the following result found in~\cite{KoSchief2} and
explained geometrically in~\cite{AD:q-red}.
\begin{Prop} \label{prop:circularity-L} A quadrilateral lattice $\vbx$ is 
circular
iff the scalars
\begin{equation}
v_i:= (T_i \vbx + \vbx)\cdot \bX_i \; , \quad i=1,\dots ,N
\end{equation}
solve the linear system~(\ref{eq:lin-X}) or, equivalently, iff the
function $|\vbx|^2$ (the square of the norm of $\vbx$) satisfies
the Laplace equation~(\ref{eq:Laplace}) of $\vbx$.
\end{Prop}

A distinguished sub-class of circular lattices corresponds to the particular
case in
which the lattice points $\vbx$ belong to the sphere of radius $R$:
$|{\vbx}|=R$.
In this case there exists, like for the symmetric reduction, an elegant
relation between point lattices and hyperplane lattices.

\begin{Prop}  \label{prop:Om-circ}
Given a system of parallel quadrilateral
lattices $\{\vbx_{(k)}\}_{k=1}^M$ and the associated matrix $\bOm$
of the system defined with respect to an orthonormal basis
$\{\vbe_k \}_{k=1}^M$, then the following properties are
equivalent.
\\ i) The matrix $\bOm /R$ is orthogonal:
\begin{equation} \label{eq:Omegaorth}
\bOm \: \bOm^T = \bOm ^T \bOm=R^2 \II, \quad \bOm^T=R^2 \bOm^{-1}.
\end{equation}
ii) The polar hyperplane ${\cal P}({\vbx}_{(k)})$ coincides with the dual
hyperplane $R^2 \vby^*_{(k)}$:
\begin{equation}
{\cal P}({\vbx}_{(k)})=R^2\vby^*_{(k)},\;\;\;k=1,..,M.
\end{equation}
iii) The quadrilateral lattices ${\vbx}_{(k)}/R,\;k=1,..,M$ form
an orthonormal basis:
\begin{equation}\label{eq:circ0}
{\vbx}_{(i)}\cdot {\vbx}_{(j)}=R^2\delta_{ij},\;\;\;i,j=1,..,M.
\end{equation}
In addition, the associated tangent vectors ${\bX}_i,\;{\bX}^*_i$,
$i=1,...,N$, are related by the following formulas
\begin{align}\label{eq:circ1}
\bX_{i}=\frac{\rho_i}{2R^2}T_i( \bOm \bX^{*T}_{i}) & =
-\frac{\rho_i}{2R^2} \bOm (T_i\bX^{*T}_{i}) , \quad i=1,\dots , N,
\\ \label{eq:circ2} T_i\bX^*_{i} & =-\frac{2}{\rho_i} \bX_i^T \bOm ,
\quad i=1,\dots , N,
\end{align}
with
\begin{equation}\label{eq:circ3}
|\bX_i|^2= \rho_i,\;\;\;T_i|{\bX}^*_i|^2= \frac{4R^2}{\rho_i}
\end{equation}
and satisfy the circularity constraint (\ref{eq:circularity1}) and
its adjoint
\begin{equation}\label{eq:circularity*}
C^{\circ *}_{ij}:={\bX}^*_i\cdot T^{-1}_i{\bX}^*_j+
{\bX}^*_j\cdot T^{-1}_j{\bX}^*_i=0.
\end{equation}
\end{Prop}

\begin{proof}
The equivalence between i) and ii) and formula (\ref{eq:circ0}) is
a straightforward consequence of the definitions of $\bOm$,
${\vbx}_{(k)}$ and $\vby^*_{(k)}$. Furthermore, the quadrilateral
lattice on a sphere is obviously circular, the circles being the
intersections of the sphere with the planes of the elementary
quadrilaterals~\cite{AD:q-red}. \\
i) $\Rightarrow$ iii). Applying $\Delta_i$ to equation 
(\ref{eq:Omegaorth}) leads
to
\begin{equation}
T_i\bX_i^{*T} \otimes \bX_i^T = -R^{-2} \bOm^T \bX_{i} \otimes
T_i(\bX^*_{i} \bOm^T), \quad i=1,..,N,
\end{equation}
which implies that
\begin{align} \label{eq:X-X*-g}
\bX_i & = \gamma_i T_i( \bOm \bX_i^{*T} ), \\ \label{eq:X*-X-g}
T_{i} \bX_i^{*}= & -\frac{1}{R^2 \gamma_i } \bX_i^T \bOm ,
\end{align}
for some $\gamma_i$. Using equation~\eqref{eq:D-Om} in
\eqref{eq:X-X*-g} one obtains
\begin{equation} \label{eq:X-X*-g2}
\bX_i = \frac{\gamma_i}{1-\gamma_i |T_i\bX_i^*|^2} \bOm
T_i\bX_i^{*T},
\end{equation}
which, together with~\eqref{eq:X*-X-g} leads to identification of
the factors $\gamma_i$
\begin{equation}
\gamma_i = \frac{2}{|T_i\bX_i^*|^2} = \frac{|\bX_i|^2}{2R^2} .
\end{equation}
Notice that equations \eqref{eq:lin-H} imply
\begin{equation}
T_iT_j\bX_i^* =  \left( 1 - (T_iQ_{ji})(T_jQ_{ij})\right)^{-1}
\left( T_i\bX_i^* + (T_i Q_{ji})T_j\bX_j^* \right) \; , \quad
i\neq j \; .
\end{equation}
Application of the shift in $j$ direction to equation
\eqref{eq:X-X*-g2} and using the above identity leads to equations
\begin{align}
T_j \gamma_i - \gamma_i (1 - (T_iQ_{ji})(T_jQ_{ij})) = 0 , \\
\gamma_i T_iQ_{ji} + \gamma_jT_jQ_{ij} + R^2 \bX_i \cdot \bX_j =
0;
\end{align}
the first of  them allows for identification $\rho_i = 2\gamma_i R^2$,
while the second gives the circularity condition.

At last, equations (\ref{eq:circ1}) and (\ref{eq:circ3}) imply the
following relation between the circularity property and its dual:
\begin{equation}
C^{\circ}_{ij}=-\frac{1}{4R^2}\frac{T_iT_j\tau}{\tau}T_iT_jC^{\circ*}_{ij},
\end{equation}
which implies that also equation (\ref{eq:circularity*}) is satisfied. The
proof of: iii) $\Rightarrow $~i)
is similar and is left to the reader.
\end{proof}
\begin{Cor}
Quadrilateral lattice in a sphere is conjugate to its own polar (with
respect to the sphere) hyperplane lattice.
\end{Cor}

In the continuous limit, equations~\eqref{eq:circularity1} become the
orthogonality conditions
\begin{equation} \label{eq:cont-ort}
\bX_i \cdot \bX_j = 0 , \qquad i\ne j
\end{equation}
and the circular lattice reduces to an orthogonal conjugate net.

\section{$D$-invariant lattice}
\label{sec:d-latt}

In this Section we introduce and discuss a basic dimensional reduction of
the quadrilateral lattice, the {\it $d$-invariant lattice}, characterized by
the invariance of a certain natural frame along the main diagonal of the
lattice.

To do so, it is convenient to put this reduction in the natural framework of
the theory of transformations of the quarilateral lattice, discussed in
great detail in~\cite{TQL}.

From a quadrilateral lattice $\bx:\ZZ^N\rightarrow\RR^M$, one can easily
construct a new quadrilateral lattice just translating $\bx$ in some
coordinate direction and combining this translation with a Combescure
transformation. If the translation takes place along the main diagonal, one
abtains the new quadrilateral lattice
\begin{equation} \label{eq:CTx}
\hat{\bx} = \calC(T\bx),
\end{equation}
where $T:=\prod_{i=1}^{N}T_i$ is the total translation along the main
diagonal and $\calC(\cdot)$ is the Combescure transformation~\cite{TQL}.
From the above definition it follows that
\begin{equation}
\D_i\hat\bx = (T_i\hat{H}_i)\hat\bX_i ,
\end{equation}
where
\begin{equation} \label{eq:hatXi}
\hat\bX_i = T\bX_i ,\qquad (\Rightarrow \hat{Q}_{ij} = TQ_{ij}),
\end{equation}
and $\hat{H}_i$ are solutions of
\begin{equation}
\D_j\hat{H}_i = (T_j\hat{H}_j)\hat{Q}_{ji}, \quad i\ne j , 
\end{equation}
different from $TH_i$.

To establish relations between quadrilateral lattices $\bx$ and $\hat\bx$,
one uses the following relations valid for generic quadrilateral lattices.
\begin{Lem} \label{Lem:TLXi}
For any subset $L=\{ i_1,\dots ,i_L \}$ of the indices
$1,2,\dots,N$, let us define the partial shift 
$T_L=\prod_{\ell =
1}^{L} T_{i_\ell} $, then 
\begin{equation} \label{eq:TLXi}
T_L \bX_i = \begin{cases}
\bX_i + \sum_{\ell \in L}(T_L Q_{i\ell}) \bX_\ell & 
\text{if \quad $i\not\in L$},\\
T_i\bX_i - (T_iQ_{ii})\bX_i + 
\sum_{\ell \in L}(T_L Q_{i\ell}) \bX_\ell & 
\text{if \quad $i \in L$},
\end{cases}
\end{equation}
where $Q_{ii}$ was defined in~\eqref{eq:Qii}.
\end{Lem}
\begin{proof}
We first prove by induction the case $i \not\in L$. For $|L|=1$ the 
statement follows 
from the linear problem~\eqref{eq:lin-X}. When $k\not\in L$ and $k\ne i$ 
and the upper part of the
formula \eqref{eq:TLXi} holds then 
\begin{multline*}
T_{L\cup\{k\}}\bX_i = T_L \left( \bX_i + (T_k Q_{ik}) \bX_k \right) = \\
\bX_i + (T_{L\cup\{k\}}Q_{ik})\bX_k + \sum_{\ell\in L} T_L \left( Q_{i\ell} +
(T_kQ_{ik})Q_{k\ell} \right) \bX_\ell,
\end{multline*}
and application of the Darboux equations~\eqref{eq:MQL-Q} concludes the
first part of the proof. Notice that applying the shifts $T_L$ and $T_k$ in
different order we obtain the following generalized Darboux equations
\begin{equation} \label{eq:TLQik}
T_L Q_{ik} = Q_{ik} + \sum_{\ell\in L}(T_L Q_{i\ell})Q_{\ell k}, \quad
i\ne k\not\in L.
\end{equation}

To show the lower part of the formula~\eqref{eq:TLXi} let us
apply the shift $T_i$ to the upper part of it obtaining
\begin{equation}
T_{L\cup\{i\}}\bX_i = T_i\bX_i + \sum_{\ell\in L}\left(
T_{L\cup\{i\}}Q_{i\ell} \right) \bX_\ell + \left( T_i \sum_{L} (T_L
Q_{i\ell})Q_{\ell i} \right) \bX_i.
\end{equation} 
It remains to prove that for a generic lattice and $i\not\in L$ 
\begin{equation}
T_LQ_{ii} =  Q_{ii} + \sum_{L} (T_L Q_{i\ell})Q_{\ell i} ,
\end{equation}
which can be done, again, by simple induction with the help of
equation~\eqref{eq:TLQik}.
\end{proof}
The quadrilateral lattice $\hat\bx$ is characterized by the following 
property.
\begin{Prop}
Let $\vbx:\ZZ^N \rightarrow \RR^M$ be a 
quadrilateral lattice and let $\hat\bx:\ZZ^N \rightarrow \RR^M$ be its
transformed quadrilateral lattice~\eqref{eq:CTx}. Then
\begin{equation} \label{eq:CTXi}
\hat\bX_i = T\bX_i= T_i\bX_i -(T_iQ_{ii})\bX_i + \sum_{\ell=1}^{N} 
\hat{Q}_{i\ell}\bX_\ell,
\end{equation}
and, consequently,
\begin{multline} \label{eq:CTQ}
\D_iQ_{ij} + \tD_j \hat{Q}_{ij} - Q_{ij} ( T_iQ_{ii} - \hat{Q}_{ii}-1) -
\hat{Q}_{ij}\left( T_j^{-1}\hat{Q}_{jj}- Q_{jj}+1\right) +\\
\sum_{\ell=1, \ell\ne i,j}^{N}\hat{Q}_{i\ell}Q_{\ell j} = 0 , \quad i\ne j.
\end{multline}
\end{Prop}
\begin{proof}
Equation~\eqref{eq:CTXi} follows from Lemma~\ref{Lem:TLXi} for
$L=\{1,\dots,N\}$ and from~\eqref{eq:hatXi}. Equation~\eqref{eq:CTQ} is the
compatibility condition of
equations~\eqref{eq:CTXi} and~\eqref{eq:lin-X}.
\end{proof}

The fixed point of transformation~\eqref{eq:CTx} (and therefore an
integrable reduction of the quadrilateral lattice) is represented by the
lattices $\bx$ which are parallel to their translations $T\bx$: $\bx =
\calC(T\bx)$ or, equivalantly, for which $T\bX_i = \bX_i$.
\begin{Def}
A quadrilateral lattice $\vbx: \ZZ^N \rightarrow \RR^M$ is {\it diagonally
invariant ($d$--invariant)} iff:
\begin{equation} \label{eq:d-inv}
T\bX_i = \bX_i \; .
\end{equation}
\end{Def}
\begin{Rem}
Equation~\eqref{eq:d-inv} implies that 
\begin{equation} \label{eq:d-inv-Q}
T Q_{ij} = Q_{ij} \; .
\end{equation}
\end{Rem}
\begin{Rem}
The $d$--invariant lattice can be described effectively by
$N-1$ parameters, since
it depends on the differences of the variables $n_i$:
\begin{equation}
\bX_i = \bX_i(n_1-n_2, n_2 - n_3 , \dots , n_{N-1} - n_N) \; .
\end{equation}
\end{Rem}
\begin{Cor}
If $\vbx$ is $d$--invariant, then $T\vbx$ is parallel to $\vbx$.
\end{Cor}

A $d$--invariant lattice is characterized by the following property.
\begin{Prop}
Let $\vbx:\ZZ^N \rightarrow \RR^M$ be a 
$d$--invariant lattice; then
\begin{equation} \label{eq:DiXi-d-inv}
\D_i\bX_i = (T_iQ_{ii})\bX_i - \sum_{\ell=1}^{N} Q_{i\ell}\bX_\ell
\end{equation}
and, consequently,
\begin{equation} \label{eq:DiQij-d-inv}
\D_iQ_{ij} + \tD_j Q_{ij} - Q_{ij} ( \D_iQ_{ii} - \tD_j Q_{jj}) +
\sum_{\ell=1, \ell\ne i,j}^{N}Q_{i\ell}Q_{\ell j} = 0 \; .
\end{equation}
\end{Prop}
\begin{proof}
Equations~\eqref{eq:DiXi-d-inv} and~\eqref{eq:DiQij-d-inv} are a
strightforward consequence of equations~\eqref{eq:CTXi} and~\eqref{eq:CTQ}
respectively.
\end{proof}
\begin{Rem}
Formula~\eqref{eq:DiXi-d-inv} implies that the $N$ dimensional $d$-invariant
lattice is effectively contained in an $N$ dimensional subspace of $\RR^M$,
therefore without loss of generality we can put in this Section $N=M$.
\end{Rem}
We present now the characterization of $d$-invariant lattices in terms
of hyperplane lattices. 
\begin{Th}
If the quadrilateral lattice $\vbx:\ZZ^N\rightarrow\RR^N$ is $d$-invariant 
then its
rotation coefficients $Q_{ij}$ are also the backward rotation coefficients
of its complementary lattice
\begin{equation} \label{eq:PQ-d-inv}
\tP^*_{ij} = Q_{ij}, \quad i\ne j = 1,\dots,N.
\end{equation}
\end{Th}
\begin{proof}
If $\vbx$ is quadrilateral, then comparison of the
formula~\eqref{eq:DiXi-d-inv} with equations~\eqref{eq:DiXi}
and~\eqref{eq:Pi} proves the statement.


\end{proof}

\section{The Egorov Lattice}
\label{sec:Eg-latt}

\begin{figure}
\begin{center}
\epsffile{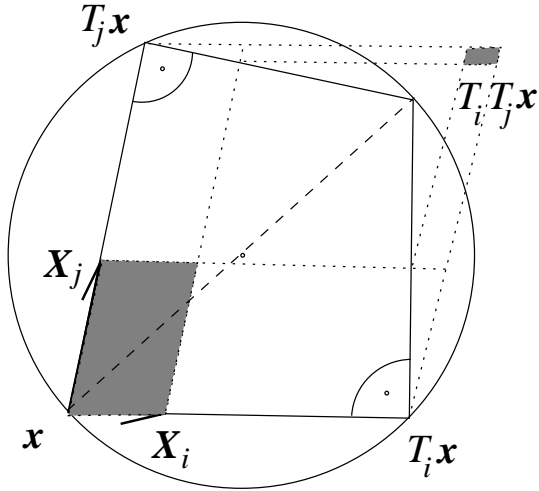}
\end{center}
\caption{Egorov lattice}
\label{fig:Egorov}
\end{figure}

\begin{Def}[\cite{Schief-priv1}]
A quadrilateral lattice is a {\it Egorov lattice} iff the
internal angles  corresponding to the
vertices $T_i\bx $ and $T_j\bx $ are right angles 
(see Figure~\ref{fig:Egorov}).
\end{Def}

Since the opposite angles of the elementary quadrilaterals of the Egorov
lattice sum up to the flat angle we have the following result.
\begin{Cor}[\cite{Schief-priv1}] 
The Egorov lattice is circular.
\end{Cor}
\begin{Rem}
The Egorov lattice constraint can be written algebraically in the form
\begin{equation} \label{eq:Eg-X}
\bX_i\cdot T_i\bX_j=0,\;\;i\ne j, 
\end{equation}
which implies the circularity condition~\eqref{eq:circularity1}.
\end{Rem}
\begin{Cor}
The line  $\langle \vbx , T_iT_j\vbx \rangle$ is a main diagonal 
of the circle defined by the points: $\vbx$, $T_i\vbx$ and $T_j\vbx$. 
\end{Cor}
\begin{Prop}
The Egorov lattice is an integrable reduction of the quadrilateral lattice.
\end{Prop}
\begin{proof}
Define functions $C^E$ by equation
\begin{equation}
C^{E}_{ij} = \bX_i\cdot T_i\bX_j ,
\end{equation}
and notice 
the following identity:
\begin{equation} \label{eq:CE-evol}
T_kC^{E}_{ij} = C^{E}_{ij} + (T_kQ_{ik})C^{E}_{kj} + 
(T_iT_kQ_{jk})C^{E}_{ik} + (T_iT_kQ_{ji})C^{E}_{ki} ,
\end{equation}
valid for a generic quadrilateral lattice.
In the case of the Egorov lattice we have $C^E=0$, 
and equation~\eqref{eq:CE-evol} shows that such constraint is admissible.
\end{proof}

In the previous sections we introduced two other basic
integrable reductions
of the quadrilateral lattice: the symmetric and the $d$-invariant
lattices. We will show that the Egorov lattice is symmetric and, for $N=M$,
$d$-invariant.
\begin{Prop}
The Egorov lattice is symmetric.
\end{Prop}
\begin{proof}
The linear problem \eqref{eq:lin-X} and the constraint~\eqref{eq:Eg-X} imply
that
\begin{equation*}
\bX_i \cdot \bX_j + (T_iQ_{ji})\bX_i \cdot \bX_i = 0 \; , \qquad i\ne j \; ,
\end{equation*}
which gives
\begin{equation} \label{eq:Eg-QX2}
(T_jQ_{ij})|\bX_j|^2=(T_iQ_{ji})|\bX_i|^2\; , \qquad i\ne j \; .
\end{equation}
Because the Egorov lattice is is circular, then $|\bX_i|^2$ can be 
identified with the potentials $\rho_i$, therefore
equation~\eqref{eq:Eg-QX2} leads to
the symmetry constraint~\eqref{eq:symm-rho}.
\end{proof}

\begin{Rem}
An equivalent form of the constraint~\eqref{eq:Eg-QX2} was used by Schief
in his derivation of the Egorov lattice from the circular
lattice~\cite{Schief-priv2}.
\end{Rem}
\begin{Rem}
The symmetry and circularity constraints are not enough to
obtain algebraically the Egorov lattice. Indeed,
consider a symmetric and circular lattice together with its tangent 
vectors $\bX_i$ and the
corresponding rotation coefficients $Q_{ij}$. 
The symmetry condition implies the
existence of a $\tau$--function (we call it $\tau^S$) such that
the potentials $\rho_i^S=T_i\tau^S/\tau^S$ satisfy
\begin{equation}
\rho_i^S T_iQ_{ji} = \rho_j^S T_j Q_{ij} .
\end{equation}
The circularity condition, in turn, implies existence of a 
$\tau$--function (we call it $\tau^C$) such that the corresponding
potentials $\rho_i^C$ are given by
\begin{equation}
\rho_i^C = |\bX_i|^2 \; .
\end{equation}
Equations \eqref{eq:scaling-f}-\eqref{eq:scaling-b2} imply that the
potentials $\rho_i^C$ and $\rho_i^S$ are connected by functions of single
variables
\begin{equation}
\rho_i^C(n) = a_i(n_i)\rho_i^S(n) \; , \quad i=1,\dots, N \; .
\end{equation}
The Egorov lattice corresponds to the distinguished case in which we have
$a_i\equiv 1$, $i=1,\dots, N$. 
\end{Rem}

\begin{Cor}
In the circular  lattice $|T_i\tbX_i| = 1/|\bX_i|$, which
implies that
the parallelogram $P(\bX_i,\bX_j)$ is 
anti-similar to the parallelogram $P(T_i\tbX_i,T_j\tbX_j)$.
In the Egorov lattice the parallelogram $P(\bX_i,\bX_j)$ is 
also anti-similar to the parallelogram $P(\D_i\bX_j,\D_j\bX_i)$.
\end{Cor}

For $N=M$ the Egorov lattice exhibits the $d$-invariance
property~\cite{Schief-priv3}.
\begin{Prop}
The Egorov lattice $\vbx:\ZZ^N\rightarrow\RR^N$ is $d$-invariant.
\end{Prop}
\begin{proof} The orthogonality conditions~\eqref{eq:Eg-X}
imply that
\begin{eqnarray}
\bX_i & \bot &\langle T_i \bX_\ell \rangle_{\ell=1, \ell\ne i}^{N} \; , \\
T\bX_i & \bot &\langle T T_\ell^{-1} \bX_\ell 
\rangle_{\ell=1, \ell\ne i}^{N} \; ,
\end{eqnarray}
where $\langle T_i \bX_\ell \rangle_{\ell=1, \ell\ne i}^{N}$ is the linear
space spanned by $\{ T_i \bX_\ell \}_{\ell=1}^{N}$, $\ell\ne i$. In
addition, the planarity of the lattice implies that these two linear
subspaces coincide; therefore $\bX_i$ and $T \bX_i$, which are orthogonal
to the same $(N-1)$ dimensional linear subspace, must be proportional:
\begin{equation} \label{eq:TX-aX}
T \bX_i = a_i \bX_i \; .
\end{equation} 
Applying $T$ to the linear system~\eqref{eq:lin-X} and
using~\eqref{eq:TX-aX}, we infer that $a_i=a_i(n_i)$ ($=1$ without loss of
genericity) and $TQ_{ij} = Q_{ij}$. 
\end{proof}

We conclude this section considering the Egorov lattice from the
point of view of the parallel system $\vbx_{(k)}$ and of its
connections with  hyperplane lattices. The results are a straightforward
consequence of the Propositions~\ref{prop:Om-symm},~\ref{prop:Om-circ}
and of the definition of the
Egorov lattice.

\begin{Prop}
Given a system of parallel quadrilateral
lattices $\{\vbx_{(k)}\}$, $k=1,..,M$ and the associated
matrix $\bOm$, then the following properties are
equivalent.

i) The matrix $\bOm /R$ is symmetric and orthogonal:
\begin{equation}\label{eq:OmegaEgorov}
\bOm^T =R^2 \bOm^{-1}=\bOm\;\;\;\Rightarrow\;\;\;\bOm^2=R^2I.
\end{equation}

ii) The polar hyperplane lattice ${\cal P}(\vbx_{(k)})$ coincides with the 
dual hyperplane lattice $R^2 \vby^*_{(k)}$ and with the adjoint hyperplane
lattice:
\begin{equation}
{\cal P}(\vbx_{(k)})=\vbx_{(k)}=R^2 {\vby}^*_{(k)},\;\;\;k=1,..,M.
\end{equation}

\end{Prop}

The continuous limit of equations \eqref{eq:MQL-Q} \eqref{eq:symm}, 
and \eqref{eq:circularity1}:
\begin{align}
\partial_i \beta_{jk} & = \beta_{ji}\beta_{ik} , & i\ne j \ne k \ne i, 
\tag{\ref{eq:Darboux}}\\
\beta_{ij} & =  \beta_{ji} , & i\ne j , \tag{\ref{eq:symm-cont}}\\
\bX_i\cdot \bX_j & = 0  , & i\ne j , \tag{\ref{eq:cont-ort}}
\end{align}
characterize submanifolds parametrized by Egorov systems of conjugate
coordinates (Egorov nets). Also, the continuous limit of
\eqref{eq:DiQij-d-inv}, together
with \eqref{eq:symm-cont} , leads to the Lam\'e equations
\begin{equation}
\partial_i\beta_{ij} + \partial_j\beta_{ji} + \sum_{\ell = 1, \ell \ne i,j}^N
\beta_{i\ell} \beta_{j\ell} = 0 
\end{equation}
which, together with equations \eqref{eq:Darboux} and \eqref{eq:symm-cont},
provide the usual characterization of a Egorov net. At last, the
$d$-invariance properties \eqref{eq:d-inv} and \eqref{eq:d-inv-Q} reduce
to
\begin{align}
\sum_{\ell = 1}^N \partial_\ell \beta_{ij} = & 0 ,\\
\sum_{\ell = 1}^N \partial_\ell \bX_{i} = & 0 ,
\end{align}
implying that $\beta_{ij} = \beta_{ij}(u_1 - u_2,\dots ,u_{N-1}-u_N)$.
For $N=3$, we recover a classical characterization of the Egorov
net~\cite{Bianchi,DarbouxOS}.

\section{$\bar\partial$ Formulations of the Reductions}
\label{sec:D-bar}

In this section we prove that the distinguished reductions of the
quadrilateral lattice discussed in the previous sections are
integrable via the $\bar\partial$ reduction method introduced in
\cite{ZakMa2} and generalized to a discrete context in \cite{DMS}.
For the sake of completeness, we first summarize in Sections~\ref{sec:db-ql}
and~\ref{sec:db-rt} the
$\bar\partial$ formulation of the quadrilateral lattice and the
main result of the $\bar\partial$ reduction theory applied to it.

The $\bar\partial$ dressing method is a very convenient tool to construct
integrable multidimensional systems, together with large classes of
solutions~\cite{ZakMa,BogdanovManakov,ZakharovSP}.
Consider the (by assumption uniquely solvable) matrix $M\times M$
${\bar\partial}$ problem
\begin{equation} \label{eq:dbar}
\partial_{\bar\lambda}\phi (\lambda ) =\partial_{\bar\lambda}\eta (\lambda )+
\int_{\CC}R(\lambda ,\lambda' )\phi (\lambda' )d\lambda'\land d\bar\lambda'
\; \; ,\;\;\;\;\;\;\lambda ,\lambda' \in {\CC},
\end{equation}
where $\partial_{\bar\lambda}={\partial /\partial\bar\lambda}$,
the given rational function $\eta (\lambda )$ (the normalization of
$\phi (\lambda )$) describes the singularities and the asymptotic behaviour
of $\phi$ in the complex plane
and $R(\lambda ,\lambda' )$ is the given  $M\times M$
matrix ${\bar\partial}$ -  datum; consider also the adjoint ${\bar\partial}$
problem:
\begin{equation} \label{eq:adjdbar}
\partial_{\bar\lambda}\phi^*(\lambda ) =-\partial_{\bar\lambda}\eta (\lambda )-
\int_{\CC}
\phi^* (\lambda' )R(\lambda' ,\lambda )d\lambda'\land d\bar\lambda'
\; \; ,\;\;\;\;\;\;\lambda ,\lambda' \in {\CC}.
\end{equation}
The above $\bar\partial$ problems imply the bilinear identity:
\begin{equation}\label{eq:bilinear}
\int_{C_{\infty}}\phi^*_2(\lambda )\phi_1 (\lambda ) d\lambda+
\int_{\CC}
[\phi^*_2 (\lambda )\partial_{\bar\lambda}\eta_1(\lambda )-
(\partial_{\bar\lambda}\eta_2(\lambda ))\phi_1 (\lambda )]
d\lambda\land d\bar\lambda =0
\end{equation}
(where $C_{\infty}$ is the circle with center at the origin and
arbitrarily large radius, and the corresponding integration is
counter-clockwise),
which involves the solutions $\phi_1$ and $\phi^*_2$ of (\ref{eq:dbar}) and
(\ref{eq:adjdbar}),
corresponding to the normalizations $\eta_1$ and $\eta_2$ respectively.

The dependence of the $M\times M$ matrices $\phi$, $\phi^*$ and $R$ on
$\bar\lambda$ and $\bar\lambda'$: $\phi = \phi(\lambda, \bar\lambda )$,
$R=R(\lambda,
\bar\lambda ,\lambda' , \bar\lambda' )$ will be omitted systematically
throughout the paper.

In the following, we shall consider
only the two basic solutions $\chi (\lambda )$
and $\chi (\lambda ,\mu ) $  of equations (\ref{eq:dbar}), corresponding
 respectively to the
``canonical normalization" $\eta =1$ and to the ``simple pole normalization"
$\eta = (\lambda -\mu )^{-1}$~\cite{GrOr1,GrOr2}, together with the corresponding
solutions of the adjoint problem (\ref{eq:adjdbar}) $\chi^*(\lambda )$ and
$\chi^*(\lambda ,\mu ) $.

\subsection{$\bar\partial$ Formulation of the Quadrilateral Lattice}
\label{sec:db-ql}

It turns out that the  MQL equations are integrable via the
$\bar\partial$ - dressing method
\cite{BoKo,DMS} and all the
geometric quantities of the lattice have a distinguished role in
this $\bar\partial$ scheme.

\begin{Prop} Let the $M\times M$ $\bar\partial$ datum $R$ depend on the
lattice variable $n=(n_1,..,n_N)\in Z^N$ in the following way
\begin{equation} \label{eq:R}
R({n};\lambda ,\lambda' )=(g({n},\lambda ))^{-1}R_0(\lambda
,\lambda' ) g({n},\lambda )
\end{equation}
\begin{equation}
g({n},\lambda )=\prod\limits_{k=1}^N[I+(\lambda -1)P_k]^{n_k},
\end{equation}
where $R_0(\lambda,\lambda' )$ is an arbitrary function of
$\lambda$ and $\lambda'$, but constant in ${n}$ and
$P_i,\;i=1,..,N$ are the usual $i$-th projection matrices:
$(P_i)_{jk}=\delta_{ij}\delta_{ik}$. Then the following results
hold.\\ 1) The matrix functions
\begin{equation}\label{eq:psidef}
\psi (\lambda ):=g({n};\lambda )\chi (\lambda ),\;\;
\psi^*(\lambda ):= \chi^*(\lambda )(g({n};\lambda ))^{-1}\;\;
\end{equation}
satisfy the following linear systems respectively:
\begin{equation} \label{eq:psisyst}
\Delta_i\psi_{kj}(\lambda )=
(T_iQ_{ji})\psi_{ki}(\lambda ),\;\;i=1,..,N,\;\;j,k=1,..,M,\;i\ne j,
\end{equation}
\begin{equation} \label{eq:adjpsisyst}
\Delta_i\psi^*_{jk}(\lambda )=
(T_i\psi^*_{ik}(\lambda ))Q_{ij},\;\;i=1,..,N,\;\;j,k=1,..,M,\;i\ne j,
\end{equation}
where $Q_{ij}$ is the $(ij)$ -- component of the matrix $Q$
defined by:
\begin{equation}\label{eq:Qdef}
Q=\displaystyle\lim_{\lambda\to\infty}{(\chi^T (\lambda )-I)}=
\displaystyle\lim_{\lambda\to\infty}{(I- \lambda ({\chi^*}^T(\lambda ))}.
\end{equation}

\noindent
2) The matrix function
\begin{equation}\label{eq:psilattdef}
\psi (\lambda ,\mu ):= g({n};\lambda )\chi (\lambda ,\mu
)(g({n};\mu ))^{-1},
\end{equation}
is connected to the canonically normalized solutions of the $\bar\partial$
problem through the equations:
\begin{equation} \label{eq:psilat}
\Delta_i\psi_{kj}(\lambda ,\mu )=
\psi_{ki}(\lambda )T_i\psi^*_{ij}(\mu ),\;\;i=1,..,N,\;\;j,k=1,..,M;
\end{equation}
furthermore the matrix function
\begin{equation}\label{eq:adjpsidef}
\psi^*(\lambda ,\mu ):= g({n};\mu )\chi^*(\lambda ,\mu
)(g({n},\lambda ))^{-1}
\end{equation}
is connected to $\psi (\lambda ,\mu )$ via:
\begin{equation}\label{eq:psipsi}
\psi^*(\mu ,\lambda )=\psi (\lambda ,\mu )
\end{equation}
and the canonically normalized solutions of the $\bar\partial$ problem can be
obtained from $\chi (\lambda ,\mu )$ via the asymptotics~\cite{BoKo-II} 
\begin{equation}\label{eq:chiliminfty}
\chi^*(\mu )=\displaystyle\lim_{\lambda\to\infty}{[\lambda\chi (\lambda ,\mu )]},
\;\;\chi (\lambda )=-\displaystyle\lim_{\mu\to\infty}{[\mu\chi (\lambda ,\mu )]}
\end{equation}
and
\begin{equation} \label{eq:chilim0}
T_i\chi_{ji}(\lambda ,0)=\chi_{ji}(\lambda )T_i\chi^*_{ii}(0),\;\;\;
\chi_{ij}(0,\mu )=-\chi_{ii}(0)T_i\chi^*_{ij}(\mu ).
\end{equation}
\end{Prop}
\begin{proof} The proof is standard, in the philosophy of the
$\bar\partial$ method.\\
1) First, after defining the ``long
derivatives"
\begin{equation*}
({\cal D}_if)(\lambda )=\Delta_if+(\lambda -1)P_iT_if,\qquad
({\cal D}^*_if)(\lambda )=-\tD_i f+(\lambda -1)(T^{-1}_i f)P_i,
\end{equation*}
one can verify that the functions
\begin{equation*}
({\cal D}_i\chi )(\lambda )P_j-\chi (\lambda )P_i(T_iQ^T)P_j,\;\;i\neq j,
\end{equation*}
\begin{equation*}
P_j({\cal D^*}_i\chi^*)(\lambda )-P_j(T^{-1}_i{Q^*}^T)P_i\chi^*(\lambda ),
\;\;i\neq j,
\end{equation*}
where $Q^*_{ij}$ is the $(ij)$ component of the matrix $Q^*$ defined by
\begin{equation}
Q^*=\displaystyle\lim_{\lambda\to\infty}{({\chi^*}^T(\lambda )-I)},
\end{equation}
solve the homogeneous version of the $\bar\partial$ problems (\ref{eq:dbar}),
(\ref{eq:adjdbar}) and go
to zero at $\lambda\to\infty$; therefore uniqueness implies the equations
\begin{equation} \label{eq:matchisyst}
({\cal D}_i\chi )(\lambda )P_j=\chi (\lambda )P_i(T_iQ^T)P_j,\;\;i\neq j,
\end{equation}
\begin{equation}
P_j({\cal D^*}_i\chi^*)(\lambda )=
P_j(T^{-1}_i{Q^*}^T)P_i\chi^*(\lambda ),\;\;i\neq j,
\end{equation}
or, equivalently, the equations
\begin{equation}
\Delta_i\psi (\lambda )P_j=\psi (\lambda )P_i(T_iQ^T)P_j,\;\;i\neq j,
\end{equation}
\begin{equation}
P_j\Delta_i\psi^*(\lambda )=
-P_j(T^{-1}_i{Q^*}^T)P_iT_i\psi^*(\lambda ),\;\;i\neq j.
\end{equation}
These two last equations, written in components, coincide with
(\ref{eq:psisyst}) and (\ref{eq:adjpsisyst}), using also the property
\begin{equation}
Q^*=-Q,
\end{equation}
which is a direct consequence of the bilinear identity
(\ref{eq:bilinear}) for $\chi (\lambda )$ and $\chi^*(\lambda )$.
At last, the $\lambda\to\infty$ limit of equation
(\ref{eq:matchisyst}) implies that the coefficients $Q_{ij}$
satisfy the MQL equations \eqref{eq:MQL-Q}.\\
2) The proof of
formulas (\ref{eq:psilat}) is conceptually similar. The function
\begin{equation}
{\cal D}_i(\chi (\lambda ,\mu )(g(\mu ))^{-1})-
\chi (\lambda )P_iT_i\varphi (\mu ),
\end{equation}
where
\begin{equation} \label{eq:varphidef}
\varphi (\mu )=\displaystyle\lim_{\lambda\to\infty}
{\lambda \chi (\lambda ,\mu )(g(\mu ))^{-1}},
\end{equation}
solves the homogeneous version of the $\bar\partial$ problem (\ref{eq:dbar})
and goes to zero at $\lambda\to\infty$; therefore uniqueness implies the
equation
\begin{equation} \label{eq:matchilat}
{\cal D}_i(\chi (\lambda ,\mu )(g(\mu ))^{-1})=
\chi (\lambda )P_iT_i\varphi (\mu ).
\end{equation}
This equation is equivalent to
\begin{equation} \label{eq:matpsilat}
\Delta_i\psi (\lambda ,\mu )=\psi (\lambda )P_iT_i\varphi (\mu ),
\end{equation}
whose component form reduces to (\ref{eq:psilat}), taking account of the
formulas
\begin{equation}\label{eq:varphipsi}
\varphi (\mu )=\psi^*(\mu ),\;\;\;\varphi^*(\mu )=-\psi(\mu ),
\end{equation}
which are obtained from the bilinear identity (\ref{eq:bilinear}) for
$\chi (\lambda ,\mu )$, $\chi^*(\lambda )$ and
$\chi^*(\lambda ,\mu )$, $\chi(\lambda )$ respectively. At last, the bilinear
identity (\ref{eq:bilinear}) for $\eta_1=(\lambda -\mu )^{-1},\;
\eta_2=(\lambda -\mu' )^{-1}$ gives $\chi^*(\mu',\mu )=\chi (\mu,\mu' )$ or,
equivalently, equation (\ref{eq:psipsi}); furthermore, equation
(\ref{eq:varphidef}),
(\ref{eq:varphipsi}) and (\ref{eq:psipsi}) lead to (\ref{eq:chiliminfty}) and
equation (\ref{eq:matchilat}),
evaluated at $\lambda =0$, gives equation (\ref{eq:chilim0}).
\end{proof}

From the solutions $\psi (\lambda ,\mu )$, $\psi (\lambda )$ and
$\psi^*(\lambda )$ of the $\bar\partial$ problem one can construct
a system $\{{\vbx}_{(k)}\},\;k=1,..,M$ of parallel quadrilateral
lattices, together with the corresponding tangent vectors and
Lam\'e coefficients, through the following matrix equations:
\begin{equation} \label{eq:Omega}
\bOm =\int_{\CC}d\lambda\wedge d\bar\lambda \int_{\CC}d\mu\wedge d\bar\mu
M(\lambda )\psi (\lambda ,\mu )M^*(\mu ),
\end{equation}
\begin{equation} \label{eq:Xi}
\bX_i=\int_{\CC}d\lambda\wedge d\bar\lambda
M(\lambda )\psi_{i}(\lambda ),\;\;
{\bX}^*_i=\int_{\CC}d\mu\wedge d\bar\mu \psi^*_{i}(\mu )M^*(\mu ),
\end{equation}
where ${\vbx}_{(i)}$ is the i-th column of matrix $\bOm$,
$\psi_{i}(\lambda )$ is the i-th column of matrix $\psi (\lambda
)$, $\psi^*_{i}(\mu )$ is the i-th row of matrix $\psi^* (\mu )$
and $M(\lambda ), \;M^*(\lambda )$ are arbitrary $M\times M$
matrices independent of ${n}$.

Finally, the evaluation of equations (\ref{eq:matchisyst}) at the
distinguished point $\lambda =0$ leads to the $\tau$ - function
representation \eqref{eq:Hir-ij}, \eqref{eq:Hir-ijk} of the MQL
lattice. Indeed, at $\lambda =0$, equations (\ref{eq:matchisyst})
read:
\begin{equation}
\Delta_i\chi_{jj}(0)=\chi_{ji}(0)T_iQ_{ji},
\end{equation}
\begin{equation} \label{eq:chi0}
\chi_{ij}(0)+\chi_{ii}(0)T_iQ_{ji}=0
\end{equation}
and imply that
\begin{equation} \label{eq:chi0tau1}
{\Delta_i\chi_{jj}(0)\over \chi_{jj}(0)}=-(T_iQ_{ji})(T_jQ_{ij}).
\end{equation}
Comparing equation (\ref{eq:chi0tau1}) with equation
\eqref{eq:rho-constr}, one is lead to the identification
\begin{equation}\label{eq:chi0tau2}
\chi_{ii}(0)=\rho_i={T_i\tau\over \tau},
\end{equation}
while equation (\ref{eq:chi0}) gives:
\begin{equation}\label{eq:chi0tau3}
\chi_{ji}(0)=-{T_j\tau_{ij}\over \tau},\;\;\;i\ne j.
\end{equation}
It is also possible to express
$\chi^*_{ii}(0)$ and  $\chi^*_{ij}(0)$ in terms of $\tau$ and $\tau_{ij}$.
To do so, we remark that the function
$\phi_2(\lambda )=T_i\chi^*(\lambda )(I+(\lambda -1)P_i)^{-1}$ satisfies
equation (\ref{eq:adjdbar}), corresponding to the forcing
$\pi \delta (\lambda )P_iT_i\chi^*(0)$. The bilinear equation
(\ref{eq:bilinear}) with this $\phi_2$
and with $\phi_1(\lambda )=\chi (\lambda )$ reduces the following equation
\begin{equation}
T_iQ^T(I-P_i)+P_i+(I-P_i)Q^T=(T_i\chi^*(0))P_i\chi (0),
\end{equation}
whose $(ii)$ and $(ij)$ - components read:
\begin{equation}
(T_i\chi^*_{ii}(0))\chi_{ii}(0)=1,\;\;(T_i\chi^*_{ji}(0))\chi_{ii}(0)=Q_{ij},
\end{equation}
implying that
\begin{equation}\label{eq:adjchi0}
\chi^*_{ii}(0)={1\over T^{-1}_i\rho_i}={T^{-1}_i\tau\over\tau},\;\;\;
\chi^*_{ji}(0)={T^{-1}_i\tau_{ij}\over \tau}.
\end{equation}

\subsection{$\bar\partial$--reduction Theory of the Quadrilateral Lattice}
\label{sec:db-rt}

The above $\bar\partial$ formulation allows one to look for
reductions of the MQL at the simpler level of the $\bar\partial$ -
datum $R$ \cite{DMS}. The particular form (\ref{eq:R}) of it
implies the following:
\begin{Prop} The following linear
constraint on the $\bar\partial$ datum $R(\lambda ,\lambda' )$:
\begin{equation} \label{eq:Rconstraint}
R^T({\lambda}^{-1},{\lambda'}^{-1})=|\lambda'|^4{\bar\lambda}^2
F(\lambda' )R(\lambda' ,\lambda )(F(\lambda ))^{-1}\; \;,
\end{equation}
gives rise to integrable reductions of the MQL. In formula
(\ref{eq:Rconstraint})
\begin{equation} \label{eq:Fdef}
F_{\pm}(\lambda )=\lambda^{-1}[A(\lambda )\pm A({\lambda}^{-1})]
\end{equation}
and $A(\lambda )$ is an arbitrary diagonal matrix.
\end{Prop}
The main implication of the constraint (\ref{eq:Rconstraint}) is
that the function $\phi^T(\lambda^{-1})F(\lambda )$  satisfies the
adjoint $\bar\partial$ problem (\ref{eq:adjdbar}) while the
function $F^{-1}(\lambda^{-1}){\phi^*}^T(\lambda^{-1})$ satisfies
the $\bar\partial$ problem (\ref{eq:dbar}):
\begin{multline}\label{eq:dbarF1}
\partial_{\bar\lambda}(\phi^T(\lambda^{-1})F(\lambda ))=
\phi^T(\lambda^{-1})\partial_{\bar\lambda}F(\lambda )+
(\partial_{\bar\lambda}\eta (\lambda^{-1}))F(\lambda )-
\\
\int_{\CC}(\phi^T({\lambda'}^{-1})F(\lambda' ))R(\lambda' ,\lambda )
d\lambda'\wedge d\bar\lambda',
\end{multline}
\begin{multline}\label{eq:dbarF2}
\partial_{\bar\lambda}(F^{-1}(\lambda^{-1}){\phi^*}^T(\lambda^{-1}))=
(\partial_{\bar\lambda}F^{-1}(\lambda^{-1})){\phi^*}^T(\lambda^{-1})-
\\ F^{-1}(\lambda^{-1})\partial_{\bar\lambda}\eta (\lambda^{-1})+
\int_{\CC} R(\lambda
,\lambda')(F^{-1}({\lambda'}^{-1}){\phi^*}^T({\lambda'}^{-1}))
d\lambda'\wedge d\bar\lambda',
\end{multline}
and these equations, through the bilinear identity (\ref{eq:bilinear}), imply
the nonlocal quadratic constraints:
\begin{multline}\label{eq:bilinearF1}
\int_{C_{\infty}}\phi^T({\lambda}^{-1})F(\lambda )\phi (\lambda )
d\lambda+ \int_{\CC} [\phi^T({\lambda}^{-1})(\partial_{\bar\lambda
} F(\lambda ))\phi (\lambda )+\\ (\partial_{\bar\lambda}\eta
(\lambda^{-1}))F(\lambda )\phi (\lambda )+ \phi^T
(\lambda^{-1})F(\lambda )\partial_{\bar\lambda}\eta (\lambda )]
d\lambda\land d\bar\lambda =0,
\end{multline}
\begin{multline}\label{eq:bilinearF2}
\int_{C_{\infty}}
\phi^*(\lambda )F^{-1}(\lambda^{-1}){\phi^*}^T(\lambda^{-1}) d\lambda+
\int_\CC [\phi^*(\lambda )(\partial_{\bar\lambda }
 F^{-1}(\lambda^{-1} )){\phi^*}^T(\lambda^{-1})-\\
(\partial_{\bar\lambda}\eta (\lambda ))F^{-1}(\lambda^{-1})
{\phi^*}^T(\lambda^{-1})-
\phi^* (\lambda )F^{-1}(\lambda^{-1})\partial_{\bar\lambda}\eta (\lambda^{-1})]
d\lambda\land d\bar\lambda =0.
\end{multline}

Therefore the constraint (\ref{eq:Rconstraint}) establishes a
nontrivial connection, whose nature depends on the particular choice of
$F(\lambda )$ (or, better, of $A(\lambda )$), between the solutions of
the $\bar\partial$ problem (\ref{eq:dbar}) and of its adjoint
(\ref{eq:adjdbar}) or, equivalently, between quadrilateral lattices and
their dual objects, the quadrilateral hyperplane lattices.
In the following we shall identify the matrix functions $A(\lambda )$
which correspond to the symmetric, circular and Egorov lattices.

\subsection{$\bar\partial$ Formulation of the Symmetric Lattice}

In this section we solve the symmetric lattice. We shall show that the
following choice:
\begin{equation} \label{eq:Fsymm}
A(\lambda )=I/2 \;\;\;\Rightarrow \;\;\;F_+(\lambda )=\lambda^{-1}I
\end{equation}
corresponds to the symmetric lattice reduction.
\begin{Prop} Let
$F(\lambda )=\lambda^{-1}I$, then the following equations hold.
\begin{equation} \label{eq:symm1}
\psi^T(\lambda ,\mu )=(\lambda\mu )^{-1}\psi^*(\mu^{-1} ,\lambda^{-1} ),
\end{equation}
\begin{equation} \label{eq:symm2}
\lambda^{-1}\psi_{ji}(\lambda^{-1})={T_i\tau\over\tau}\psi^*_{ij}(\lambda ),
\end{equation}
\begin{equation}\label{eq:symm3}
\chi^T(0)=\chi (0)
\end{equation}
and equations (\ref{eq:Omega}), (\ref{eq:Xi}) allow to construct a system of
symmetric lattices provided that
\begin{equation}\label{eq:symmM}
M^*(\lambda )=\lambda |\lambda |^{-4}M^T(\lambda ).
\end{equation}
\end{Prop}
\begin{proof}
We use the same strategy of the previous
$\bar\partial$ proofs. Comparing equation (\ref{eq:dbarF1}) with
equation (\ref{eq:adjdbar}) for $\eta =(\lambda -\mu )^{-1}$, one
obtains
\begin{equation}
\chi^T(\lambda^{-1},\mu^{-1})=\lambda\mu \chi^*(\lambda ,\mu )
\end{equation}
or, equivalently, (\ref{eq:symm1}), using equations (\ref{eq:psilat}),
(\ref{eq:psipsi}).
Furthermore one can verify that $T_i\chi^*(\lambda )(I-(\lambda^{-1}-1)P_i)$
satisfies the $\bar\partial$ equation (\ref{eq:dbar}), for
$\eta =T_i\chi^*(0)P_i$. Therefore, taking account of the $\lambda$ - large
asymptotics, one obtains the equation
\begin{equation}
T_i\chi^*(\lambda )(I-(\lambda^{-1}-1)P_i)-\lambda^{-1}(T_i\chi^*(0))P_i
\chi^T(\lambda^{-1})=(I-P_i)\chi^*(\lambda ),
\end{equation}
whose $(ij)$ component gives (\ref{eq:symm2}), using equations
(\ref{eq:psidef}) and (\ref{eq:adjchi0}). At last, equation
(\ref{eq:bilinearF1}) for $\eta =1$ gives directly
(\ref{eq:symm3}), which can be immediately identified with the
symmetry constraint (46), using equations (\ref{eq:chi0tau2}),
(\ref{eq:chi0tau3}) and \eqref{eq:tau}. Furthermore, equations
(\ref{eq:symm1}),(\ref{eq:symm2}) imply equations
\eqref{eq:Omegasymm} and \eqref{eq:symmX}, provided that one uses
(\ref{eq:symmM}).
\end{proof}

\medskip

\subsection{$\bar\partial$ Formulation of the Circular Lattice}

It was shown in \cite{DMS} that the following choice
\begin{equation}
A(\lambda )=(\lambda -1)^{-1}I\;\;\Rightarrow \;\;F_-(\lambda )=
{\lambda +1\over \lambda (\lambda -1)}I
\end{equation}
corresponds to the circular lattice reduction.

\begin{Prop} Let $F(\lambda )={\lambda +1\over \lambda (\lambda -1)}I$,
then the following equations hold:
\begin{align} \label{eq:circ1db}
\chi (0)+\chi^T(0)&=2\chi^T(1)\chi (1), \\ \label{eq:adjcirc1db}
\chi^*(0)+{\chi^*}^T(0)& =2\chi^*(-1){\chi^*}^T(-1),
\\ \label{eq:circ2db}
{\lambda +1\over \lambda (1-\lambda
)}\chi^T(\lambda^{-1},\mu^{-1})& = {\mu (\mu +1)\over 1-\mu }\chi
(\mu ,\lambda )+ \chi^T(1,\mu^{-1})\chi (1,\lambda ),
\\ \label{eq:adjcirc2db}
{\lambda -1\over \lambda (1+\lambda
)}\chi^T(\mu^{-1},\lambda^{-1}) & = {\mu (\mu -1)\over 1+\mu }\chi
(\lambda ,\mu )+ \chi (\lambda ,-1)\chi^T(\mu^{-1},-1),
\\ \label{eq:circ3db}
4\chi^T(1,-1)\chi (1,-1) & =I.
\end{align}
\end{Prop}
\begin{proof} Equations (\ref{eq:bilinearF1}) and
(\ref{eq:bilinearF2}) for $\eta =1$ give respectively
(\ref{eq:circ1db}) and (\ref{eq:adjcirc1db}). 
Consider equation (\ref{eq:dbarF1}) for $\eta
=(\lambda -\mu )^{-1}$, then equation (\ref{eq:circ2db}) follows
from the fact that its RHS satisfies equation
(\ref{eq:bilinearF1}) as well. Analogous considerations lead to
equation (\ref{eq:adjcirc2db}). At last equation
(\ref{eq:circ2db}), evaluated at $\lambda =\mu =-1$, gives the
orthogonality condition (\ref{eq:circ3db}).
\end{proof}

To show that the above formulas give rise to a circular lattice,
consider the following identification:
\begin{equation}\label{eq:circlat}
{\vbx}_{(i)}=(\psi_{1i}(1,\mu ),..,\psi_{Mi}(1,\mu ))^T,
\end{equation}
\begin{equation}\label{eq:db-circ-id-X}
{\bX}_i=(\psi_{1i}(1),..,\psi_{Mi}(1))^T,\;\;H_{i(n)}=
\psi^*_{in}(\mu),
\end{equation}
\begin{equation} 
{\bX}^*_i=(\psi^*_{i1}(-1),..,\psi^*_{iM}(-1)).
\end{equation}
Because of equations (\ref{eq:chi0tau2}), the diagonal part of
(\ref{eq:circ1db}) leads to
\begin{equation}\label{eq:chi0X}
\chi_{ii}(0)=\rho_i=  |{\bX}_i|^2,
\end{equation}
while the off-diagonal part gives the circularity constraint
\eqref{eq:circularity1}. Evaluating equation (\ref{eq:circ2db}) at
$\mu =0$ and using equation (\ref{eq:chiliminfty}), one obtains
\begin{equation}
{\lambda +1\over \lambda (\lambda -1)}\chi^T(\lambda^{-1})=
\chi (0 ,\lambda )-2\chi^T(1)\chi (1,\lambda ),
\end{equation}
which, using equations (\ref{eq:chilim0}) and (\ref{eq:chi0X}), can be written
in the following form:
\begin{equation}
-{\lambda +1\over \lambda (\lambda -1)}\psi (\lambda^{-1})=
({\vbx}_{(k)}+T_i{\vbx}_{(k)})\cdot \bX_k, \;\;\;k=1,..,M,
\end{equation}
which is the $\bar\partial$ formulation of the first point of
Proposition \ref{prop:circularity-L}. If, instead, we choose $\mu
=\lambda^{-1}$ we obtain
\begin{equation}
{\lambda +1\over \lambda (1 - \lambda)}[\psi (\lambda^{-1},\lambda )+
\psi^T (\lambda^{-1},\lambda )]=\psi^T(1,\lambda )\psi (1,\lambda )
\end{equation}
which, through the identification (\ref{eq:circlat}), leads to
\begin{equation}\label{eq:xjxk}
{\lambda +1\over \lambda (1- \lambda)}[\psi_{jk}(\lambda^{-1},\lambda )+
\psi_{kj}(\lambda^{-1},\lambda )]={\vbx}_{(j)}\cdot {\vbx}_{(k)}.
\end{equation}
This formula states that the scalar product of the two parallel
lattices ${\vbx}_{(j)}$, ${\vbx}_{(k)},\;j\ne k$, such that
\begin{equation}
\Delta_i{\vbx}_{(j)}=(T_iH_{i(j)})\bX_i,\;\;H_{i(n)}=\psi^*_{in}(\mu
),
\end{equation}
is equal to the sum of two scalar solutions of the Laplace
equation \eqref{eq:Laplace}, \eqref{def:A-H}, corresponding
respectively to the Lam\'e coefficients $H_{i(j)},\;H_{i(k)}$. If
$j=k$, equation (\ref{eq:xjxk}) reduces to
\begin{equation}
|{\vbx}_{(j)}|^2= {2(\lambda +1)\over \lambda (1 - \lambda)}
\psi_{jj}(\lambda^{-1},\lambda ),
\end{equation}
which is the $\bar\partial$ formulation of the second point of
Proposition \ref{prop:circularity-L}. Equation
(\ref{eq:adjcirc1db}) expresses the circularity condition
\eqref{eq:circularity*} for hyperplane lattices through the
identification (\ref{eq:circlat}) and equation (\ref{eq:circ3db})
is the $\bar\partial$ formulation of equation
\eqref{eq:Omegaorth}, through the identification:
\begin{equation}
\bOm =\psi (1,-1),\;\;R=2.
\end{equation}
In this case, both systems $\{{\vbx}_{(k)}\}$ and
$\{{\vbx}^*_{(k)}\}$ are circular. We finally remark that
equations (\ref{eq:circ2db}), (\ref{eq:adjcirc2db}) contain all
the other circular constraints for a suitable choice of $\lambda$
and $\mu$.

\subsection{$\bar\partial$ Formulation of the $D$-invariant Lattice}
The $d$-invariance lattice, a distinguished reduction of the quadrilateral
lattice, corresponds to the following distributional $\bar\partial$ datum
\begin{equation} \label{eq:R-din}
R(\lambda,\lambda') = \frac{i}{2}\delta(\lambda-\lambda')R(\lambda)
\end{equation}
and is solved by the {\emph local} $\bar\partial$ problem
\begin{align} \label{eq:d-inv-xi}
\partial_{\bar\lambda}\chi(\lambda) &= \chi(\lambda)R(\lambda)\\
\label{eq:d-inv-R}
T_iR(\lambda) &= \left[ 1 + (\lambda -1)P_i\right]R(\lambda)
\left[ 1 + (\lambda -1)P_i\right]^{-1}.
\end{align}
If $N=M$, from equation~\eqref{eq:d-inv-R} it follows the 
invariance property
\begin{equation}
TR(\lambda) = R(\lambda)
\end{equation}
which implies that
\begin{equation}
T\chi(\lambda) = \chi(\lambda).
\end{equation}
Consequently:
\begin{align}
T\psi(\lambda) &= \lambda\psi(\lambda),\\
TQ&= Q,\\
T\rho_i &=\rho_i,
\end{align}
and, taking $\lambda=1$, we obtain formulae~\eqref{eq:d-inv},~\eqref{eq:d-inv-Q}.

\subsection{$\bar\partial$ Formulation of the Egorov Lattice}

The Egorov lattice is circular and symmetric; therefore the corresponding 
constraints are
satisfied simultaneously; i.e., 
\begin{equation}\label{eq:egorR1}
|\lambda'|^{-4}{\bar\lambda}^{-2}R^T({\lambda'}^{-1},{\lambda}^{-1})=
\lambda R(\lambda ,\lambda' ){\lambda'}^{-1}=
({\lambda +1\over \lambda (\lambda -1)})^{-1}
R(\lambda ,\lambda' ){\lambda' +1\over \lambda' (\lambda' -1)}
\end{equation}
This implies the equation
\begin{equation}\label{eq:egorR2}
{2\lambda (\lambda -\lambda') \over
\lambda'(\lambda'-1)(\lambda +1)} R(\lambda ,\lambda' )=0,
\end{equation}
which admits the distributional solution~\eqref{eq:R-din}.
Therefore the $\bar\partial$ formulation of the Egorov lattice is
given in terms of the {\em local} $\bar\partial$ problem
\eqref{eq:d-inv-xi}, \eqref{eq:d-inv-R} in which the $\bar\partial$-datum
satisfies the constraint
\begin{equation} \label{eq:db-eg-const}
R^T(\lambda^{-1})={\bar\lambda}^2 R(\lambda ).
\end{equation}

Because of this locality, the corresponding
$\bar\partial$--reduction theory of Section~\ref{sec:db-rt} simplifies
considerably. 

The constraint \eqref{eq:db-eg-const} implies that
$\chi^T(\lambda^{-1})$ is a solution of the adjoint $\bar\partial$ problem:
\begin{equation}
\partial_{\bar\lambda}\chi^* (\lambda )= - \chi^* (\lambda )R(\lambda ),
\end{equation}
and the corresponding quadratic constraint:
\begin{equation}
\partial_{\bar\lambda}\left(\chi^T(\lambda^{-1}) \chi(\lambda) \right) = 0,
\end{equation}
together with the asymptotics $\lim_{\lambda\to\infty}\chi^T(\lambda^{-1})
\chi(\lambda) = \chi^T(0)$, imply that
\begin{equation}
\chi^T(\lambda^{-1}) \chi(\lambda) = \chi^T(0) .
\end{equation}
Evaluating this constraint at $\lambda=1$ and using the identifications
\eqref{eq:db-circ-id-X}, its diagonal part gives \eqref{eq:chi0X}, while its
off-diagonal part gives the Egorov constraint \eqref{eq:Eg-X}.

\section*{Acknowledgements}
The main part of the paper was written during the stay of A.~D. 
at the Rome Section of the {\it Istituto Nazionale di Fisica Nucleare}
as post-doctoral fellow. A.~D. was partially supported
by the Polish Committee of Scientific Research (KBN) under the Grant No.~ 2 
P03B 143 15.

\end{document}